\documentclass[sigconf,screen]{acmart}

\renewcommand\footnotetextcopyrightpermission[1]{}

\AtBeginDocument{%
  \providecommand\BibTeX{{%
    \normalfont B\kern-0.5em{\scshape i\kern-0.25em b}\kern-0.8em\TeX}}}
    
\usepackage{booktabs} 
\usepackage{subcaption} 
\usepackage{enumitem}

\newcommand{\todo}[1]{\textcolor{black}{#1}}

\acmConference[ESEC/FSE 2020]{The 28th ACM Joint European Software Engineering Conference and Symposium on the Foundations of Software Engineering}{8 - 13 November, 2020}{Sacramento, California, United States}

\begin{document}

\title{An Empirical Study of Usages, Updates and Risks of Third-Party Libraries in Java Projects}

\author{Ying Wang}
\affiliation{
\institution{Fudan University}
\country{China}
\vspace{-1pt}
}

\author{Bihuan Chen}
\affiliation{
\institution{Fudan University}
\country{China}
\vspace{-1pt}
}

\author{Kaifeng Huang}
\affiliation{
\institution{Fudan University}
\country{China}
\vspace{-1pt}
}

\author{Bowen Shi}
\affiliation{
\institution{Fudan University}
\country{China}
\vspace{-1pt}
}

\author{Congying Xu}
\affiliation{
\institution{Fudan University}
\country{China}
\vspace{-1pt}
}

\author{Xin Peng}
\affiliation{
\institution{Fudan University}
\country{China}
\vspace{-1pt}
}

\author{Yang Liu}
\affiliation{
\institution{Nanyang Technological University}
\country{Singapore}
\vspace{-1pt}
}

\author{Yijian Wu}
\affiliation{
\institution{Fudan University}
\country{China}
}


\begin{abstract}
Third-party libraries are a central building block to develop software systems. However, outdated third-party libraries are commonly used, and developers are usually less aware of the potential risks. Therefore, a quantitative and holistic study on usages, updates and risks of third-party libraries can provide practical insights to improve the ecosystem sustainably. In this paper, we conduct such a study in the Java ecosystem. Specifically, we conduct a \textit{library usage analysis} (e.g., usage intensity and outdatedness) and a \textit{library update analysis} (e.g., update intensity and delay) using \todo{806} open-source projects. The two analyses aim to quantify usage and update practices holistically from the perspective of both open-source projects and third-party libraries. Then, we conduct a \textit{library risk analysis} (e.g., potential risk and developer response) in terms of bugs with \todo{15} popularly-used third-party libraries. This analysis aims to quantify the potential risk of using outdated libraries and the developer response to the risk. Our findings from the three analyses provide practical insights to developers and researchers on problems and potential solutions in maintaining third-party libraries (e.g., smart alerting and automated updating of outdated libraries). To demonstrate the usefulness of our findings, we propose a bug-driven alerting system for assisting developers to make confident decisions in updating third-party library versions. We have released our dataset to foster valuable applications and improve the ecosystem.


\end{abstract}

%
%
%


\maketitle


\section{Introduction}\label{sec:intro}

Third-party libraries allow developers to reuse common functionalities instead of reinventing the wheel, and thus substantially improve the productivity of developers. In contrast to the benefits third-party libraries bring to software development, developers incur tremendous costs in software maintenance to keep third-party libraries up-to-date. Old third-party library versions contain various bugs that cause crashes or increase the attack surface of software systems, while new third-party library versions fix bugs, refactor code or add features, which may break library APIs~\cite{Kim2011EIR,Bogart2016BAC}. Therefore, the usages and updates of third-party libraries are a double-edged sword, demanding a thorough assessment of benefits and costs. 

To understand the usage and update practices of third-party libraries in the Java ecosystem, some studies have explored the usage trend and popularity of third-party library versions~\cite{mileva2009mining, kula2017exploratory}, classes and interfaces~\cite{Mileva2010, hora2015apiwave} or APIs~\cite{lammel2011large, de2013multi}, and some studies have analyzed the reasons for updating or not updating third-party libraries~\cite{bavota2015apache} and the usages of latest/outdated third-party libraries~\cite{kula2015trusting, kula2018developers}. Similar studies also exist in the npm~\cite{wittern2016look} and Android~\cite{li2016investigation, Derr2017KMU} ecosystem. However, existing studies fail to \textit{quantitatively} characterize the usage and update practices of third-party libraries in software systems and to \textit{holistically} characterize the practices from the perspective of all involved parties (i.e., software systems and third-party libraries). Thus, it lacks concrete and comprehensive evidences on how intensively software systems use and update third-party libraries and how intensively third-party libraries are used and updated across software systems. This situation hides problems in maintaining third-party libraries, and hinders practical solutions. 

Moreover, despite some recent advances from both academics (e.g., \cite{cadariu2015tracking, PontaPS2018beyond}) and industries (e.g., Black Duck~\cite{blackduck}, SourceClear~\cite{sourceclear}, Snyk~\cite{snyk} and Greenkeeper~\cite{greenkeeper}) on alerting developers to new third-party library versions, many software systems still use outdated third-party libraries, as reported from different ecosystems (e.g., Java~\cite{kula2015trusting, kula2018developers}, Android~\cite{Derr2017KMU, Salza2018DUT} and npm~\cite{lauinger2017thou, zerouali2018empirical}). These alerting systems are mostly security-driven; i.e., they only notify developers about security bugs in old third-party library versions, but leave developers unaware of potential risks of non-security bugs and potential efforts to update third-party libraries. This situation hinders their wide adoption.

To improve on such situations sustainably, it is important to quantitatively and holistically characterize usages, updates and risks of third-party libraries in software systems. This paper makes such a characterization in the Java ecosystem by answering three research questions:

\begin{itemize}[leftmargin=*]
\item \textbf{RQ1: Library Usage Analysis.} What is the usage intensity and usage outdatedness of third-party libraries?

\item \textbf{RQ2: Library Update Analysis.} What is the update intensity and update delay of third-party libraries?

\item \textbf{RQ3: Library Risk Analysis.} What is the potential risk and developer response of outdated third-party libraries?
\end{itemize}

Using \todo{806} Java open-source projects and \todo{13,565} third-party libraries, we conduct \textit{library usage analysis} and \textit{library update analysis} through a quantitative study from the perspective of both open-source projects and third-party libraries. Using \todo{15} popular third-party libraries in \todo{806} projects, we perform \textit{library risk analysis} by a quantitative study and a human study. 

Through these analyses, we aim to quantify usages, updates and risks of third-party libraries and provide useful findings to developers and researchers. For example, 77.4\% of libraries are used in only one project. 60.0\% of libraries have at most 2\% of their APIs called across projects. 38.0\% of projects use libraries that are more than 10 versions away from the latest versions. 54.9\% of projects leave more than half of the library dependencies never updated. 50.5\% of projects have an update delay of more than 60 days. 31.6\% of library version releases of the 15 popular libraries have severe bugs. Developers need convincing and fine-grained information to decide whether to update outdated, buggy library versions.

Our findings help to uncover problems in maintaining third-party libraries for open-source project and third-party library developers, quantify the significance or severity of these problems to raise attention in the ecosystem, and enable follow-up research to address these problems; e.g., smart alerting and automated updating of outdated libraries, usage-driven library evolution, and automated library bug triggering. 

To demonstrate the usefulness of our findings, we propose a bug-driven alerting system to provide multiple fine-grained information to assist developers to make confident decisions about third-party library version updates. Our preliminary results show that \todo{98.0\%} of the \todo{433} open-source projects that use buggy third-party library versions can be safe. For the \todo{9} unsafe open-source projects, \todo{we quantify the risk, impact and integration effort at a fine-grained method level to provide an informative view.}

In summary, this paper makes the following contributions:
\begin{itemize}[leftmargin=*]
\item We conducted large-scale analyses to quantitatively and holistically characterize the usages, updates and risks of third-party libraries in Java open-source projects.
\item We provided practical implications to developers and researchers, released our dataset to foster applications and improve the ecosystem, and proposed a prototype system to demonstrate the usefulness of our findings.
\end{itemize}

The rest of this paper is organized as follows. Sec.~\ref{sec:design} introduces our study design. Sec.~\ref{sec:usage}, \ref{sec:update} and \ref{sec:risk} respectively report the findings for our library usage analysis, library update analysis and library risk analysis. Sec.~\ref{sec:implication} discusses the implications and applications of our findings and the threats to our study. Sec.~\ref{sec:related} reviews the related work before Sec.~\ref{sec:conclusion} draws the conclusions.


\section{Empirical Study Methodology}\label{sec:design}

In this section, we first introduce the design of our empirical study, and then present our corpus selection process.

\subsection{Study Design}

Our study aims to characterize usages, updates and risks of third-party libraries in Java open-source projects. To this end, we propose the three RQs as introduced in Sec.~\ref{sec:intro}. For the ease of presentation, hereafter we refer to \textit{third-party library} as \textit{library} and \textit{Java open-source project} as \textit{project} if there is no ambiguity. Before introducing the design of RQs, we define library terms to avoid confusion. A \textit{library version} is a library with the version number. A \textit{library version release} is a library version with the release information (e.g., release date). A \textit{library dependency} is a library version declared as a dependency in a project.

Our library usage analysis (RQ1) analyzes the currently-used libraries in projects. We first measure how intensively a project depends on libraries (i.e., usage intensity from the perspective of projects) and how intensively a library is adopted across projects (i.e., usage intensity from the perspective of libraries). It aims to  quantify the significance of libraries in project development and the impact of evolving libraries. Then, we measure how far the adopted library versions are from the latest versions (i.e., usage outdatedness) from the perspective of projects and libraries. It aims to quantify the commonness of adopting outdated libraries in projects and motivate the necessity of RQ2. 

Our library update analysis (RQ2) investigates the historical library version updates in projects. We first measure how intensively a project updates library versions (i.e., update intensity from the perspective of projects) and how intensively a library's versions are updated across projects (i.e., update intensity from the perspective of libraries). It aims to quantify the practices of updating library versions. Then, we measure how long library version updates lag behind library version releases (i.e., update delay) from the perspective of projects and libraries. It aims to quantify the project developers' reaction speed to new library version releases and motivate the necessity of RQ3.

Our library risk analysis (RQ3) investigates the severe bugs in popular libraries. We first measure how many severe bugs exist in a library version release (i.e., potential risk). It aims to quantify the potential risk of adopting outdated library versions and delaying library version updates. Then, we explore how project developers respond to buggy library versions in their projects (i.e., developer response). It aims to characterize the project developers' reaction to buggy library versions and their requirements of an alerting system for outdated library versions.


\begin{figure*}[!t]
\centering
\begin{subfigure}[b]{0.5\textwidth}
\centering
\includegraphics[width=0.99\textwidth]{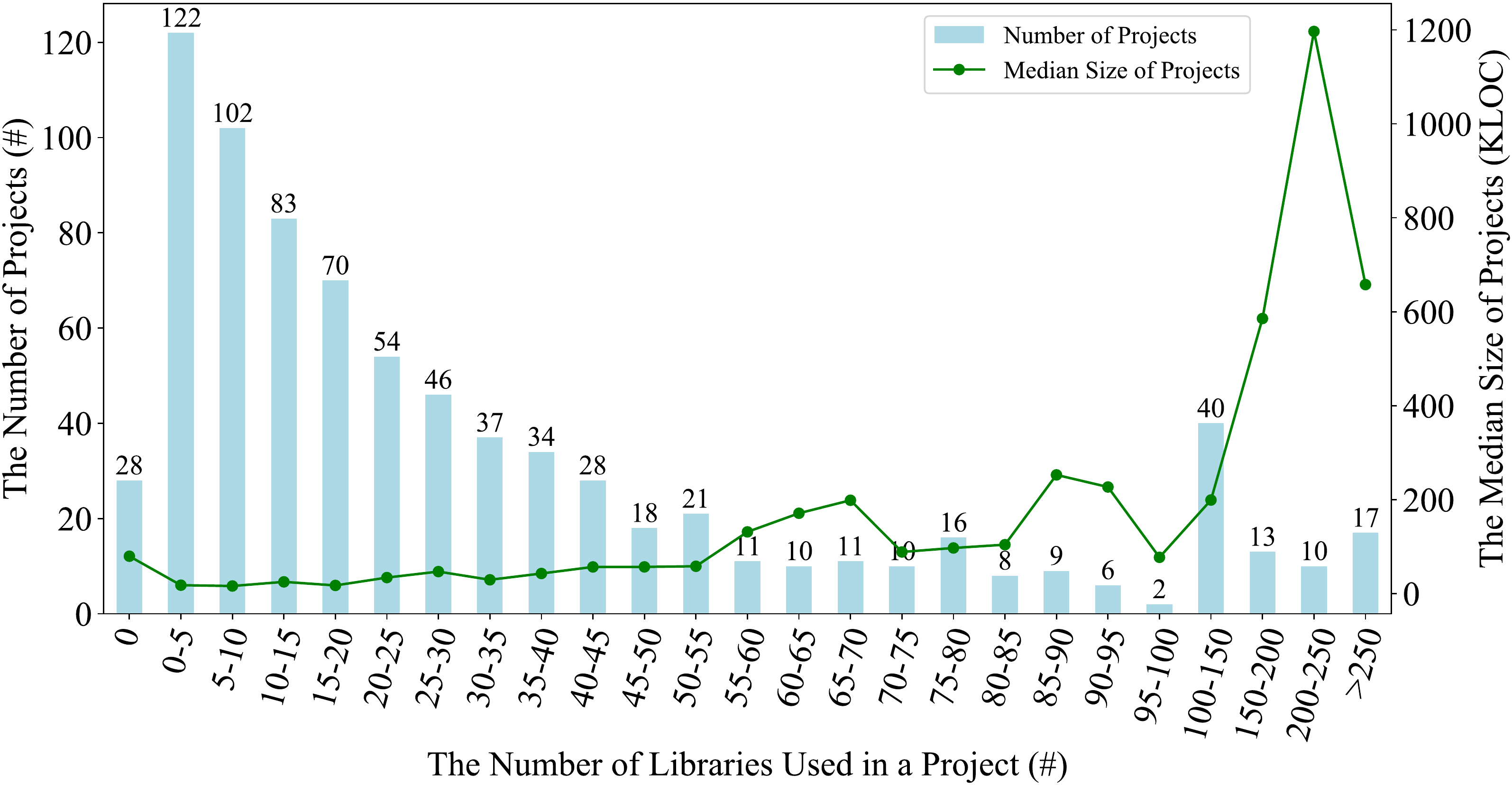}
\caption{Library-Level Usage Intensity across Projects}
\label{fig:LUIP}
\end{subfigure}~~~
\begin{subfigure}[b]{0.445\textwidth}
\centering
\includegraphics[width=0.99\textwidth]{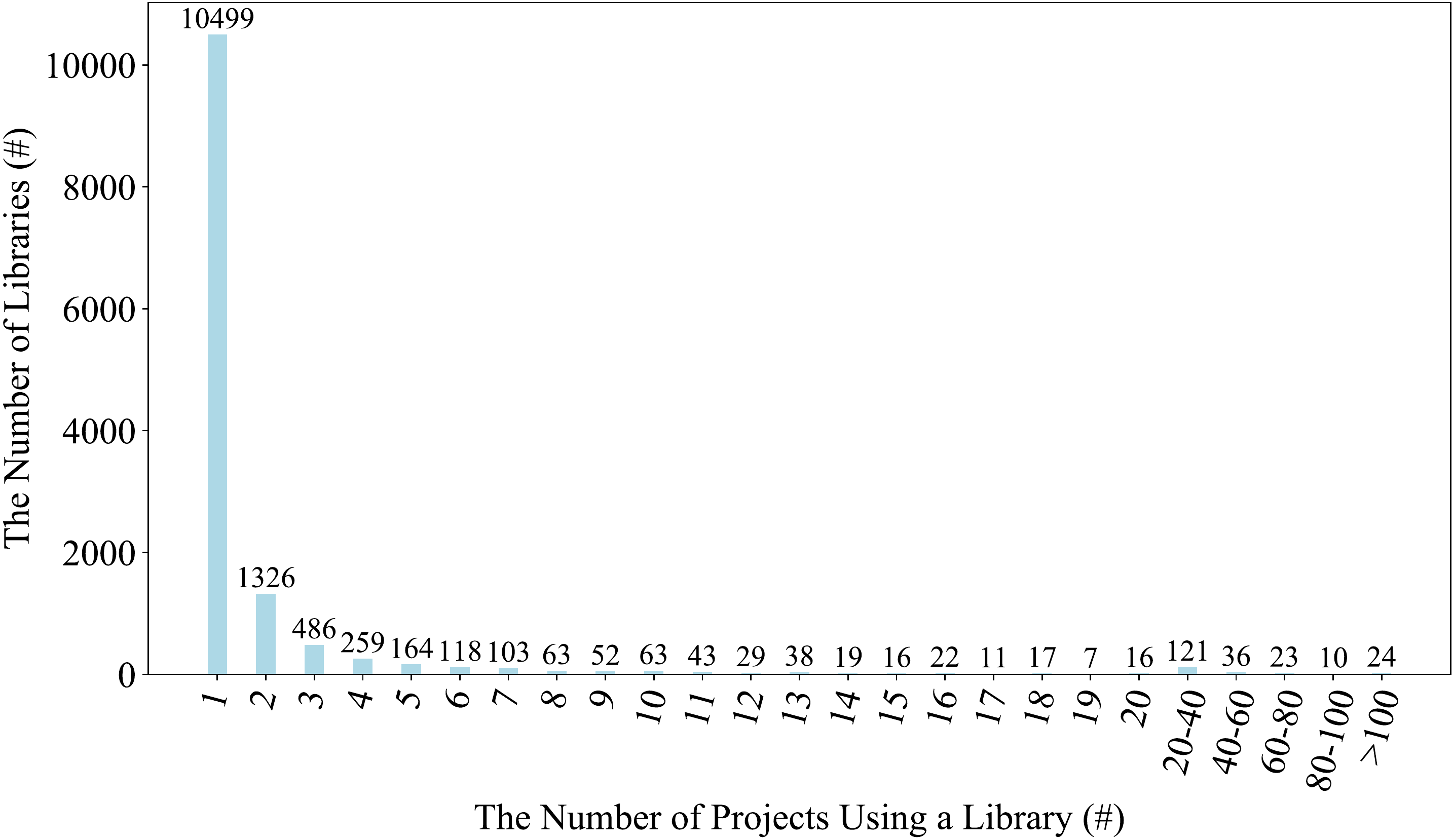}
\caption{Library-Level Usage Intensity across Libraries}
\label{fig:LUIL}
\end{subfigure}
\caption{Distributions of Library-Level Usage Intensity across Projects and Libraries}
\end{figure*}

\begin{figure*}[!t]
\centering
\begin{subfigure}[b]{0.44\textwidth}
\centering
\includegraphics[width=0.99\textwidth]{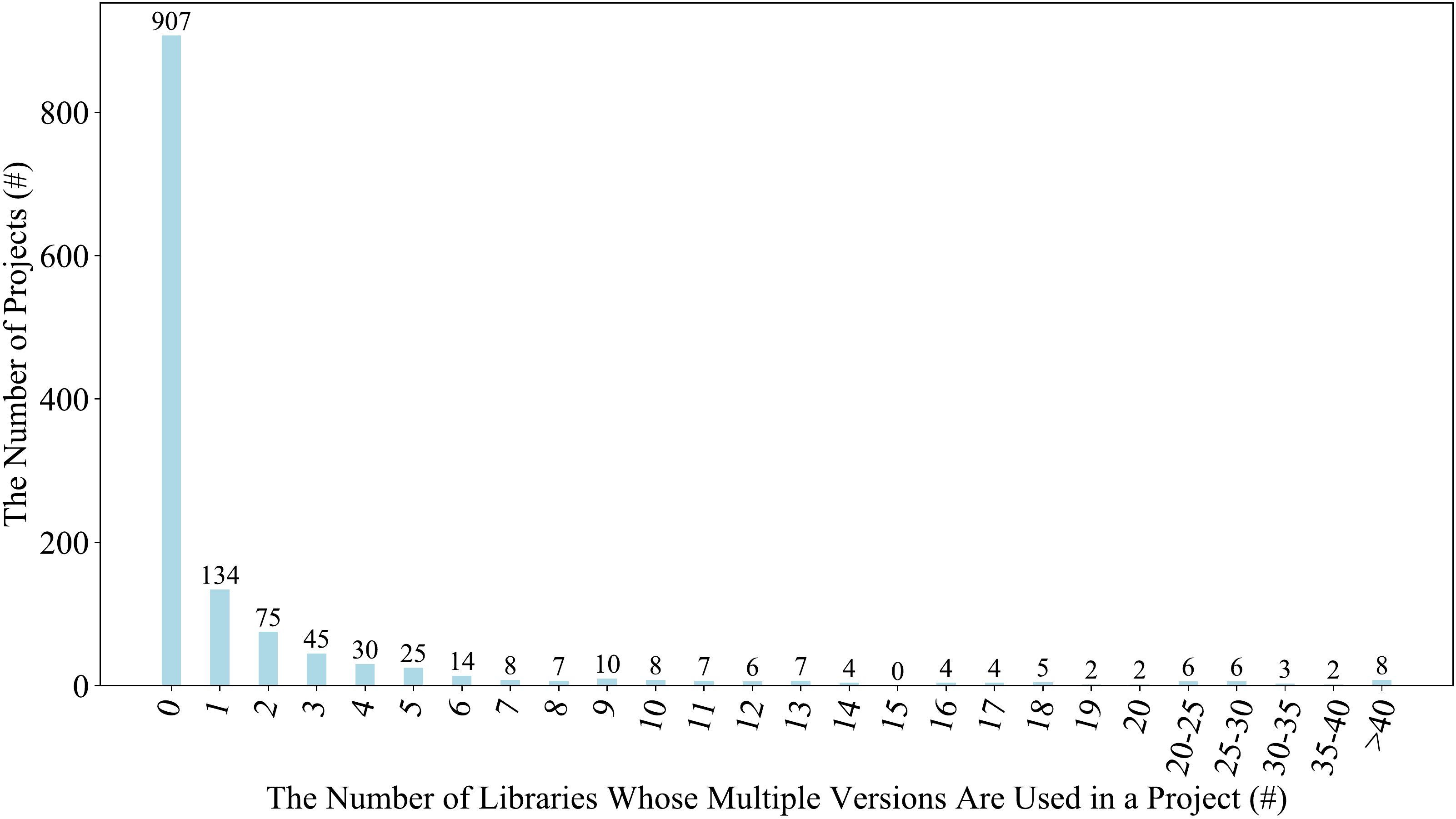}
\caption{Number of Libraries of Multiple Versions}
\label{fig:MVL}
\end{subfigure}~~~
\begin{subfigure}[b]{0.445\textwidth}
\centering
\includegraphics[width=0.99\textwidth]{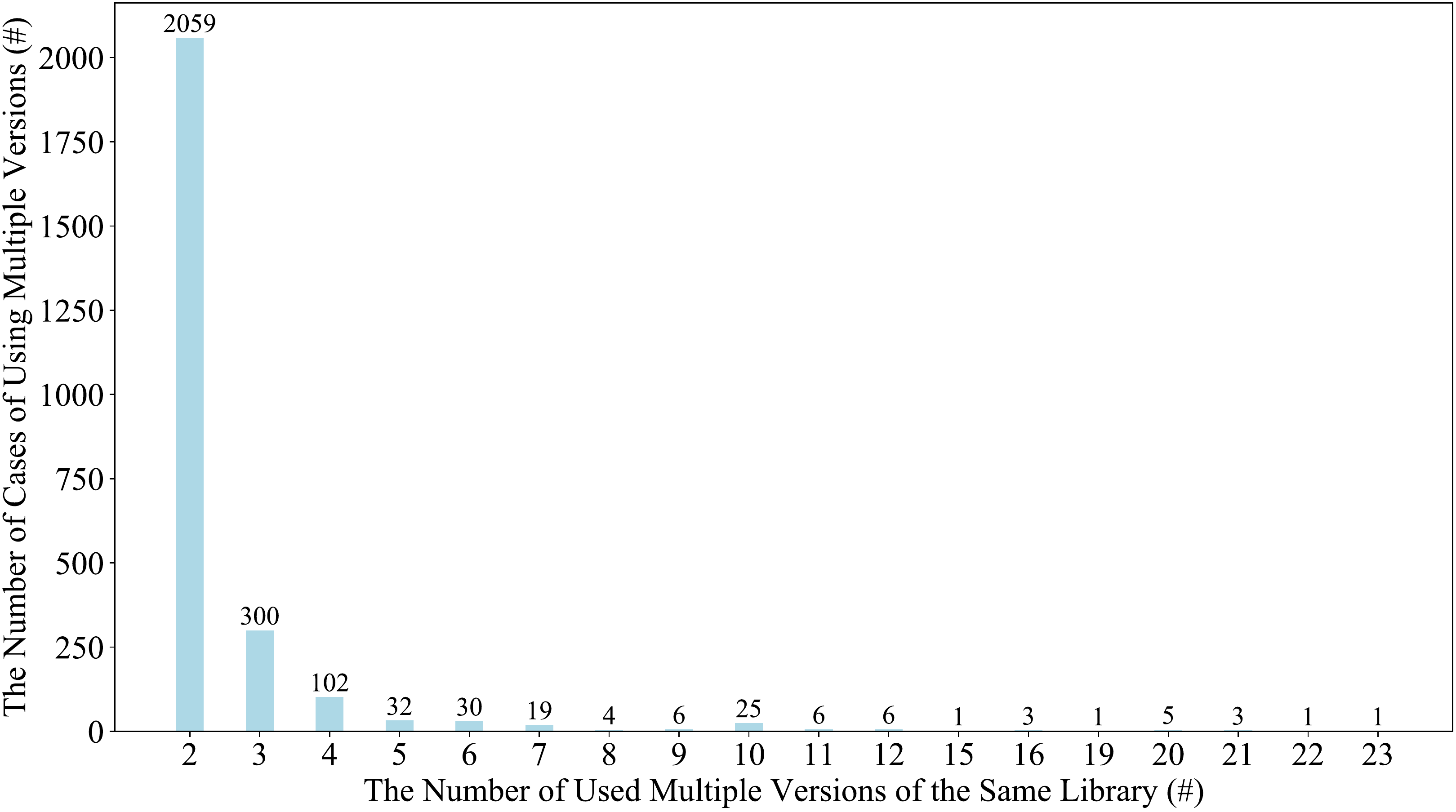}
\caption{Number of Used Multiple Versions}
\label{fig:MVV}
\end{subfigure}
\caption{Severity of Using Multiple Versions of the Same Library}
\end{figure*}

\subsection{Corpus Selection}


We conducted this study on a corpus of Java open-source projects selected from GitHub. We focused on Java because it is widely-used and thus our findings can be beneficial to wide audiences. Specifically, we first selected Java projects that had more than 200 stars to ensure the project quality, which resulted in an initial set of \todo{2,216} projects. Of these projects, we selected projects that used Maven or Gradle as the automated build tool in order to easily extract the declared library dependencies in projects, which restricted our selection to a set of \todo{1,828} projects. To obtain representative library usage and update data, we picked active and well-maintained projects that had commit in the last three months. Finally, we had a set of \todo{806} projects, denoted as $\mathcal{P}$. We crawled their repositories and commits from GitHub as of \todo{July 22, 2018} on a desktop with 2.29 GHz Intel Core i5 CPU and 8 GB RAM. Using this corpus, we conducted library crawling and library analyses on the same desktop, which took a total of four months.


\section{Library Usage Analysis}\label{sec:usage}

To analyze library usages, we develop \textit{lib-extractor} to extract library dependencies from each project's configuration file (i.e., \textit{pom.xml} and \textit{build.gradle} for Maven and Gradle projects) in a commit. One project can have multiple configuration files, especially when the project contains multiple modules. Typically, for Maven projects, \textit{lib-extractor} extracts a library dependency via parsing three fields: \textit{groupId}, \textit{artifactId} and \textit{version}; and for Gradle projects, it extracts a library dependency by parsing similar fields: \textit{group}, \textit{name} and \textit{version}. Notice that Maven and Gradle support various mechanisms (e.g., inheritance and variable expansion) to declare library dependencies and we support them in \textit{lib-extractor}. 

A library dependency $d$ is denoted as a 4-tuple $\langle p, f,$ $com, v \rangle$, where $p$ and $f$ denote the project and configuration file where $d$ is declared (here $p.date$ denotes the date when $p$'s repository is crawled), $com$ denotes the commit where $d$ is extracted (here $com.date$ denotes the date when $com$ is submitted), and $v$ denotes the library version declared in $d$. $v$ is denoted as a 2-tuple $\langle l, ver \rangle$, where $l$ denotes a library, and $ver$ denotes $l$'s version number. $l$ is denoted as a 2-tuple $\langle group, name \rangle$, where $group$ denotes $l$'s organization, and $name$ denotes $l$'s name. 

As RQ1 targeted libraries currently used in projects, we used \textit{lib-extractor} to the latest commit of each project in $\mathcal{P}$, and extracted \todo{164,470} library dependencies, denoted as $\mathcal{D}_{us}$. From $\mathcal{D}_{us}$, we identified \todo{24,205} library versions, denoted as $\mathcal{V}_{us}$ (i.e., $\mathcal{V}_{us} = \{d.v | d \in \mathcal{D}_{us}\}$). From $\mathcal{V}_{us}$, we identified \todo{13,565} libraries, denoted as $\mathcal{L}_{us}$ (i.e., $\mathcal{L}_{us} = \{v.l | v \in \mathcal{V}_{us}\}$).

\subsection{Library-Level Usage Intensity}\label{sec:usageIntensityL}

\textbf{Definition.} We define usage intensity at a coarse-grained (i.e., library) level from the perspective of a project and a library: $usi_p^1$, the number of libraries that are adopted in a project $p$, and $usi_l^1$, the number of projects that adopt a library $l$. Using $\mathcal{P}$, $\mathcal{L}_{us}$ and $\mathcal{D}_{us}$, we compute $usi_p^1$ and $usi_l^1$ by Eq. \ref{eq:usi1}.

\begin{small}
\begin{equation}
\begin{aligned}
\forall p \in \mathcal{P}, usi_p^1 = &|\{d.v.l | d \in \mathcal{D}_{us} \wedge d.p = p\}| \\
\forall l \in \mathcal{L}_{us}, usi_l^1 = &|\{d.p | d \in \mathcal{D}_{us} \wedge d.v.l = l\}|
\end{aligned}\label{eq:usi1}
\end{equation}
\end{small}

\begin{figure*}[!t]
\centering
\begin{subfigure}[b]{0.443\textwidth}
\centering
\includegraphics[width=0.99\textwidth]{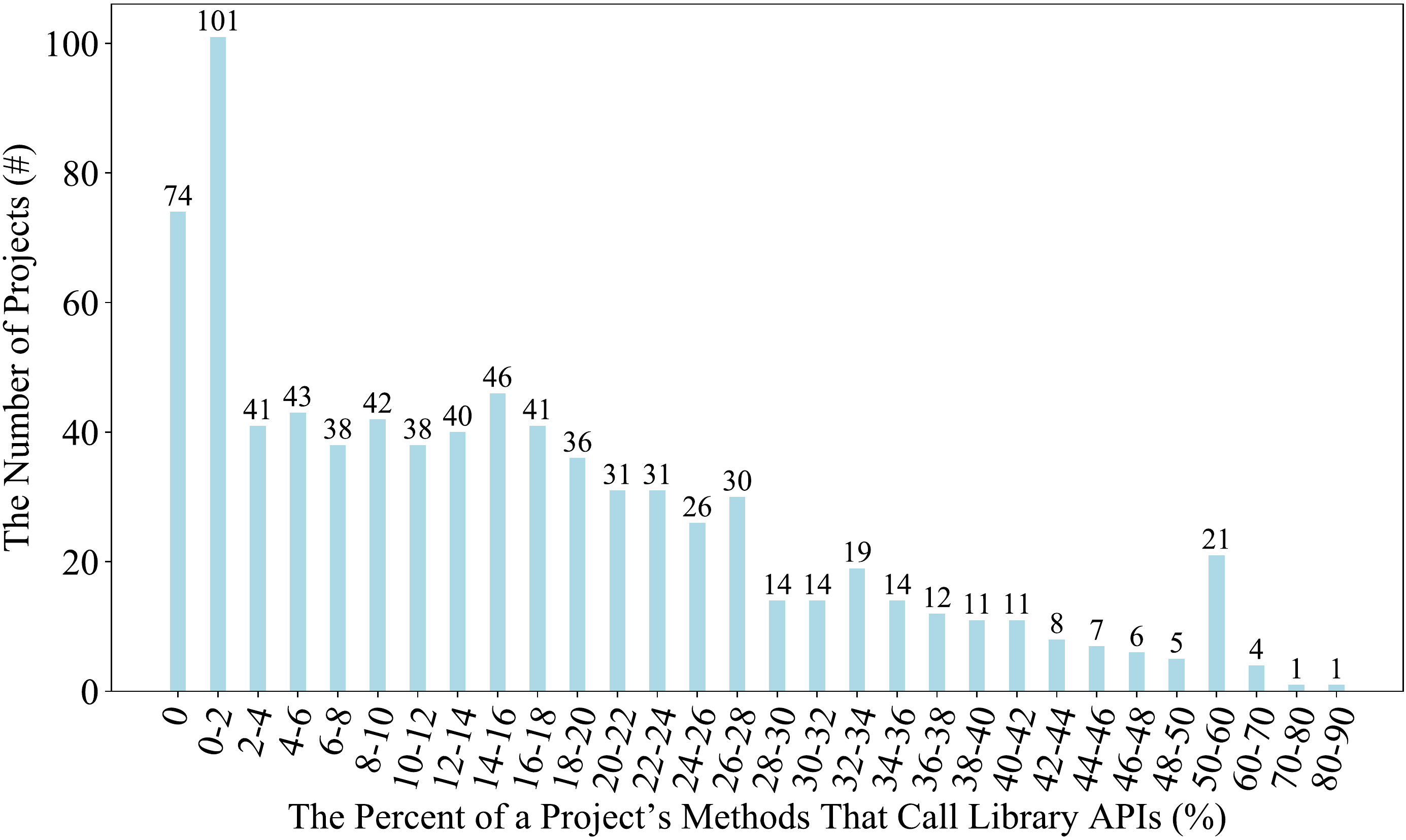}
\caption{Method-Level Usage Intensity across Projects}
\label{fig:MUIP}
\end{subfigure}~~~
\begin{subfigure}[b]{0.45\textwidth}
\centering
\includegraphics[width=0.99\textwidth]{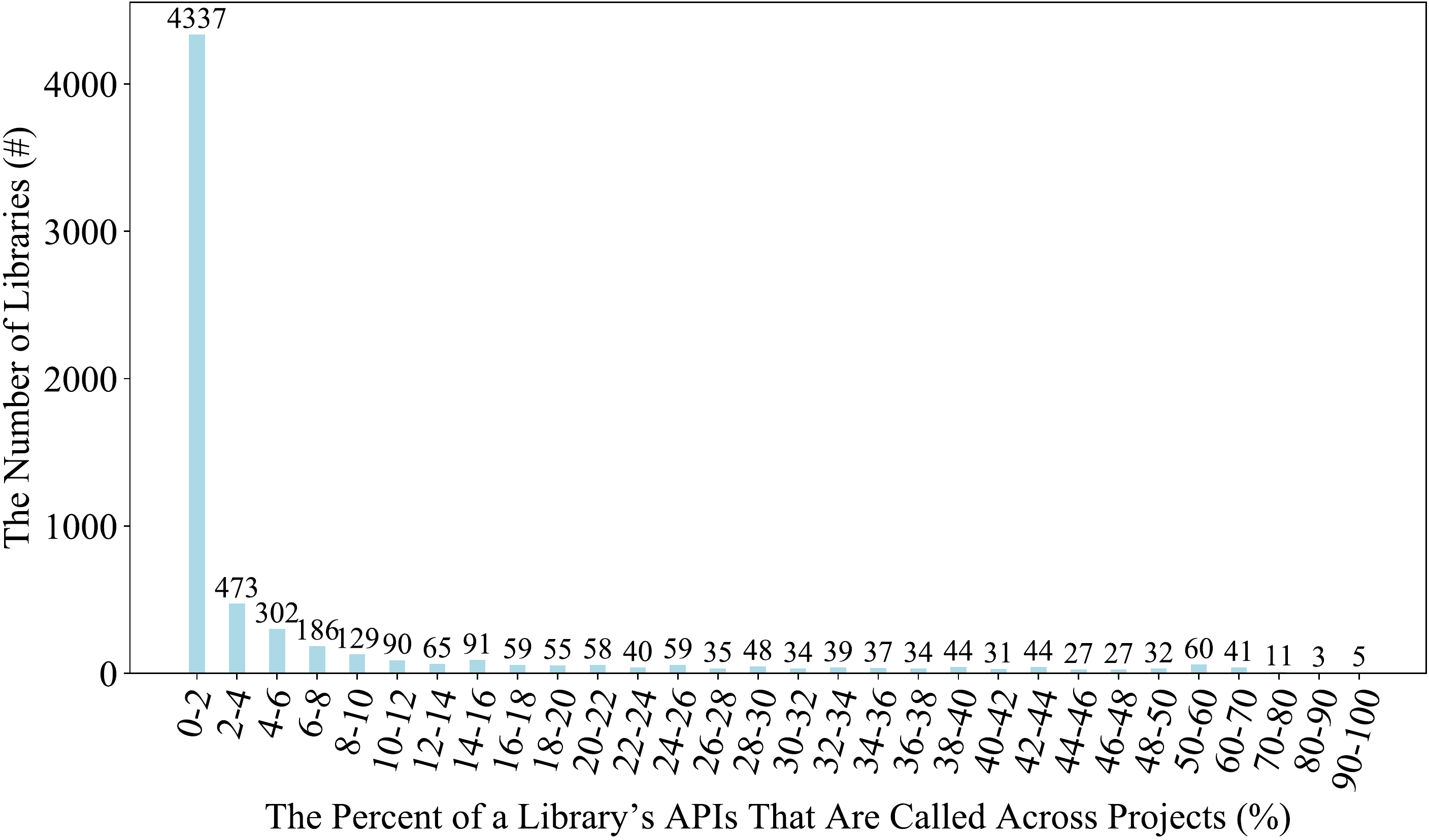}
\caption{Method-Level Usage Intensity across Libraries}
\label{fig:MUIL}
\end{subfigure}%
\caption{Distributions of Method-Level Usage Intensity across Projects and Libraries}
\end{figure*}

\textbf{Basic Findings.} Using $usi_p^1$ and $usi_l^1$, we show distributions of usage intensity across projects and libraries in Fig. \ref{fig:LUIP} and \ref{fig:LUIL}, where the $y$-axis respectively represents the number of projects and libraries whose usage intensity falls into a range. On the one hand, 28 (3.5\%) projects do not adopt libraries; and 224 (27.8\%) projects use at most 10 libraries. 401 (49.8\%), 184 (22.8\%) and 80 (9.9\%) projects respectively adopt more than 20, 50 and 100 libraries. On the other hand, 10,499 (77.4\%) libraries are used in only one project; and only 432 (3.2\%) and 214 (1.6\%) libraries are respectively adopted in more than 10 and 20 projects. These findings demonstrate that i) project development has a moderate dependency on libraries, and ii) only a very small portion of libraries are widely adopted across projects.

\textbf{In-Depth Findings.} As different modules in a project can declare library dependencies separately to suit different needs and release schedules and work around incompatibility issues \cite{Wang2018DCM}, we investigate the adopted library versions, and find that multiple versions of the same library are used in different modules of a project. Therefore, we analyze the severity of using multiple versions from two perspectives: the number of libraries whose multiple versions are used in a project, and the number of used multiple versions of the same library. The results are reported in Fig. \ref{fig:MVL} and \ref{fig:MVV}, respectively. Overall, 300 (37.2\%) projects adopt multiple versions of the same library in different modules. In detail, 84 (10.4\%) and 57 (7.1\%) projects contain one and two libraries whose multiple versions are used, respectively; and 84 (10.4\%) projects involve more than five libraries whose multiple versions are used. Moreover, among the 2,032 cases of using multiple versions of the same library, 1,600 (78.7\%) and 233 (11.5\%) cases involve two and three versions of the same library, respectively; and 95 (4.7\%) cases use more than five versions of the same library. These findings reveal the commonness of adopting multiple versions of the same library, which might increase library maintenance cost, or even lead to dependency conflicts when modules are inter-dependent.

Intuitively, the number of libraries adopted in a project is correlated to the size of the project as large projects can be complex and have high needs for libraries. To explore this conjecture, we measure the project size in thousands of lines of code (KLOC), and report the median size of the projects in each bar in Fig. \ref{fig:LUIP}. Overall, projects using a large number of libraries are larger in size than projects using a small number of libraries. Such a difference is statistically significant (i.e., $p =$ 0.00003 in one-way ANOVA test \cite{Howell2012}). In an extreme case, \textit{Apache Camel} adopts 1,290 libraries, and its size is 1,326 KLOC. This finding implies that library maintenance is a non-trivial task in large projects as it is difficult for project developers to have a clear vision of the libraries used in the large code base.

To understand the library characteristics affecting usage intensity, we analyze library categories. Unfortunately, only one-fifth of the libraries in $\mathcal{L}_{us}$ (in Fig. \ref{fig:LUIL}) have their category specified on library repositories. Hence, we directly choose the 50 most popular libraries (as the number of popular libraries is relatively small), but sample 371 libraries from 10,499 libraries used by one project. The sample size allows the generalization of our results at a confidence level of 95\% and a margin of error of 5\%, computed by a sample size calculator \cite{calculator}. Then, three of the authors follow an open coding procedure \cite{khandkar2009open} to manually categorize libraries based on the category list from Maven. Among the 50 most popular libraries, 33 are general-purpose libraries (e.g., 12 testing libraries, 10 utility libraries and 8 logging libraries) that provide functions of general interest, while 17 are domain-specific libraries (e.g., 6 web libraries and 4 database libraries) that belong to popular domains. Among the 371 unpopular libraries, only 53 share the same 17 (out of 24) categories to those 50 popular ones, and are mostly functionally-similar libraries; and the others mostly provide specific function (e.g., 129 Eclipse plugin libraries) that is only of interest for certain projects across 89 categories.

\subsection{Method-Level Usage Intensity}\label{sec:usageIntensityM}

\textbf{Definition.} We further define usage intensity at a fine-grained (i.e., method) level from the perspective of a project and a library: $usi_p^2$, the percent of a project $p$'s methods that call library APIs, and $usi_l^2$, the percent of a library $l$'s APIs that are called across projects. To compute $usi_p^2$ and $usi_l^2$, we need to extract library APIs, project methods, and API calls in project methods.

Therefore, we first crawled the jar file of each library version $v \in \mathcal{V}_{us}$ from library repositories (e.g., Maven and Sonatype) declared in configuration files. Of the \todo{24,205} library versions in $\mathcal{V}_{us}$, we crawled jar files for \todo{16,384} library versions, denoted as $\tilde{\mathcal{V}}_{us}$, but failed for \todo{7,821} (32.3\%) library versions. The main reason is that 76.1\% of such library versions are snapshot versions\footnote{a.k.a. changing versions whose features are under active development but are allowed for project developers to integrate before stable versions are released.} whose jar files are no longer available at library repositories; and the other reasons are very old library versions that are no longer available and private libraries that we do not have permissions to access. From $\tilde{\mathcal{V}}_{us}$, we identified \todo{7,229} libraries, denoted as $\tilde{\mathcal{L}}_{us}$.

\begin{figure}[!t]
\centering
\includegraphics[width=0.45\textwidth]{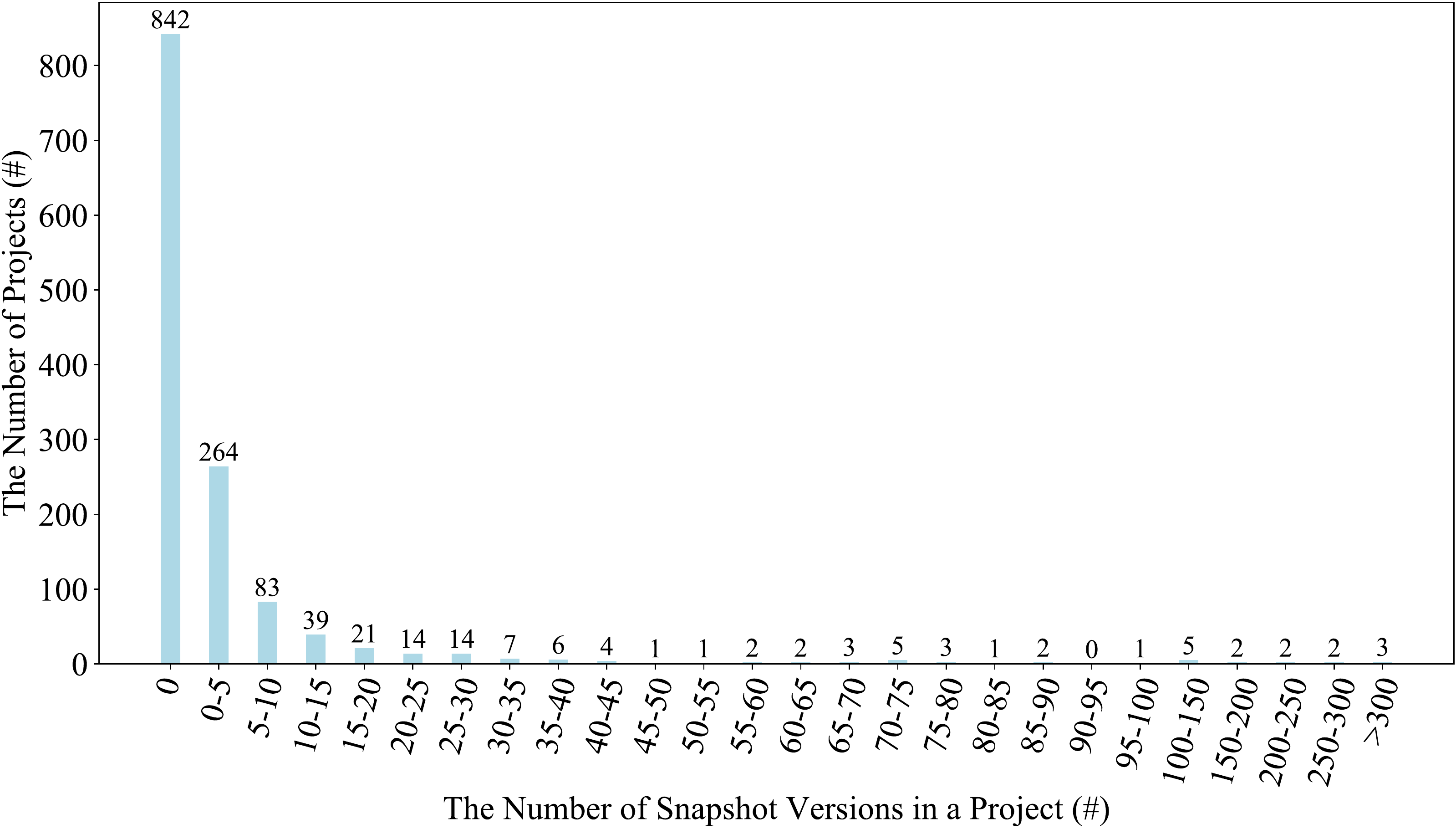}
\caption{Severity of Using Snapshot Versions}\label{fig:snapshot}
\end{figure}

Then, we applied Soot \cite{vallee2010soot} on the crawled jar files for $\tilde{\mathcal{V}}_{us}$ to extract library APIs, denoted as $\mathcal{A}$. Each library API $a \in \mathcal{A}$ is denoted as a 2-tuple $\langle v, api \rangle$, where $v \in \tilde{\mathcal{V}}_{us}$ denotes a library version, and $api$ denotes a library API. Here, we conservatively treat public methods and fields in public classes as library APIs. Next, we applied JavaParser \cite{smith2017javaparser} on the crawled project repositories and jar files of the used library versions to extract project methods, denoted as $\mathcal{M}$, and API calls in project methods, denoted as $\mathcal{C}$. Each project method $m \in \mathcal{M}$ is denoted as a 2-tuple $\langle p, method \rangle$, where $p$ denotes a project, and $method$ denotes a method in $p$. Each API call $c \in \mathcal{C}$ is denoted as a 2-tuple $\langle a, m \rangle$, where $a \in \mathcal{A}$ denotes a library API, and $m \in \mathcal{M}$ denotes the project method where $a$ is called. 

Using $\mathcal{P}$, $\tilde{\mathcal{V}}_{us}$, $\tilde{\mathcal{L}}_{us}$, $\mathcal{A}$, $\mathcal{M}$ and $\mathcal{C}$, we compute $usi_p^2$ and $usi_l^2$ by Eq. \ref{eq:usi2}. Notice that $\mathcal{V}_l = \{v | v \in \tilde{\mathcal{V}}_{us} \wedge v.l = l\}$ denotes $l$'s used versions, and $usi_l^2$ takes their maximum usage intensity.

\begin{small}
\begin{equation}
\begin{aligned}
\forall p \in \mathcal{P}, usi_p^2 =& \frac{|\{c.m | c \in \mathcal{C} \wedge c.m.p = p\}|}{|\{m | m \in \mathcal{M} \wedge m.p = p\}|} \\
\forall l \in \tilde{\mathcal{L}}_{us}, usi_l^2 =& \max_{v \in \mathcal{V}_l}{\frac{|\{c.a | c \in \mathcal{C} \wedge c.a.v = v\}|}{|\{a | a \in \mathcal{A} \wedge a.v = v\}|}} 
\end{aligned}\label{eq:usi2}
\end{equation}
\end{small}

\textbf{Basic Findings.} Using $usi_p^2$ and $usi_l^2$, we show distributions of usage intensity across projects and libraries in Fig. \ref{fig:MUIP} and \ref{fig:MUIL}. On the one hand, 74 (9.2\%) projects do not call library APIs: 28 projects do not use libraries, 5 projects use library versions that are unavailable, and 41 projects only use the resource files in jar files. 265 (32.9\%) projects have at most 10\% of methods calling library APIs. 266 (33.0\%) and 64 (7.9\%) projects respectively have more than 20\% and 40\% of methods that call library APIs. On the other hand, 4,337 (60.0\%) libraries have at most 2\% of their APIs called across projects; and only 281 (3.9\%) libraries have more than 40\% of their APIs called across projects. Notice that 733 libraries do not have class files, but only have resource files in the jar files (e.g., \textit{Angular} only contains web assets), and hence are not included in Fig. \ref{fig:MUIL}. These findings indicate that i) project developers need to make moderate effort on maintaining libraries (e.g., updating library versions and migrating to new libraries with similar functions), and ii) only a very small part of library APIs in most libraries are called across projects.

\textbf{In-Depth Findings.} As snapshot versions are the main reason for unavailable jar files, we explore the used library versions and find that 344 (42.7\%) projects use snapshot versions. Therefore, we measure the severity of snapshot versions in terms of the number of snapshot versions used in a project. The result is reported in Fig. \ref{fig:snapshot}. Overall, 7,345 (30.3\%) of the library versions in $\mathcal{V}_{us}$ are snapshot versions, and 5,951 (81.0\%) of them are no longer available at library repositories. As shown in Fig. \ref{fig:snapshot}, 161 (20.0\%) and 183 (22.7\%) projects adopt at most and more than five snapshot versions, respectively. Such findings reveal the common usages of snapshot versions, which may increase maintenance cost and risk of incompatible APIs.

We follow the same procedure as in Sec. \ref{sec:usageIntensityL} to analyze the categories of 50 libraries directly taken from the tail of Fig. \ref{fig:MUIL} and 353 libraries statistically sampled from the first bar in Fig. \ref{fig:MUIL}. The 50 libraries mostly provide very specific but relatively simple functions (e.g., XML processing, stream processing, and rule engine), and thus a relatively large portion of their APIs are used. On the opposite, the 353 libraries are mostly provide very specific but complex functions (e.g., distributed processing, and mocking), require platform integration (e.g., Eclipse and Maven plugins), or contain a wide range of related functions (e.g., Java specification), and thus a large part of APIs are not designed for client usages, or not widely used due to different function needs.

\subsection{Usage Outdatedness}\label{sec:usageOutdatedness}

\begin{figure*}[!t]
\centering
\begin{subfigure}[b]{0.42\textwidth}
\centering
\includegraphics[width=0.99\textwidth]{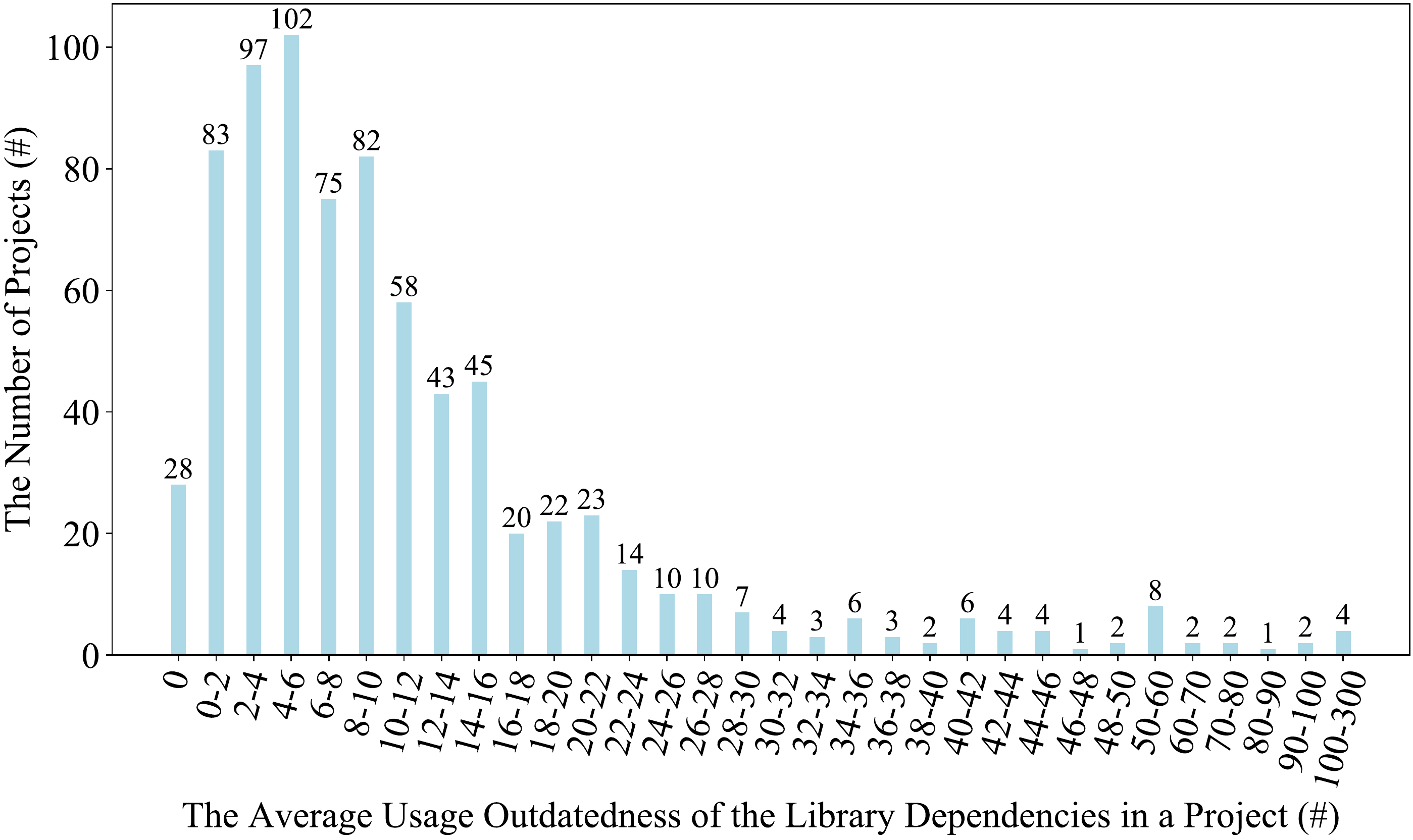}
\caption{Usage Outdatedness across Projects}
\label{fig:UOP}
\end{subfigure}~~~
\begin{subfigure}[b]{0.485\textwidth}
\centering
\includegraphics[width=0.99\textwidth]{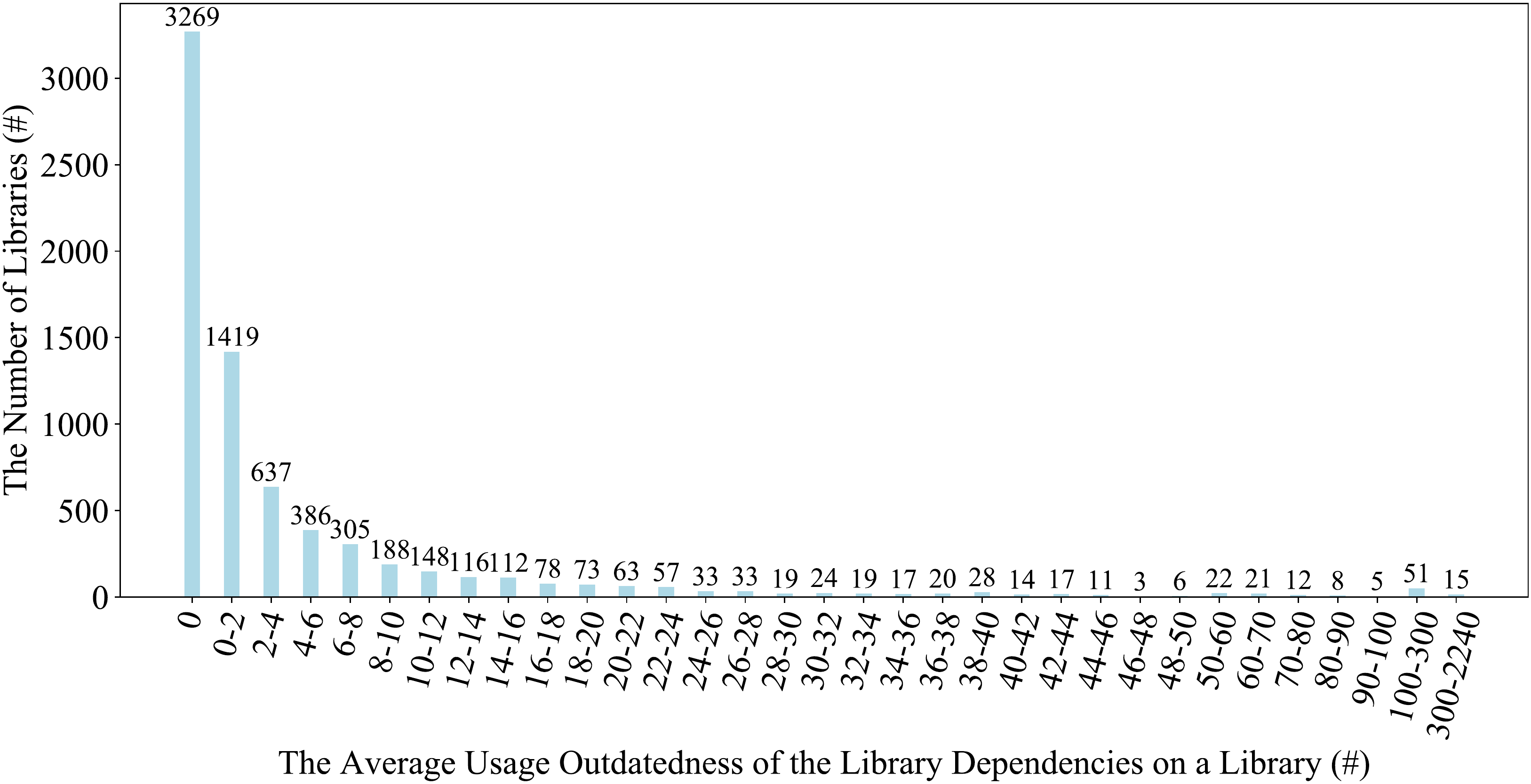}
\caption{Usage Outdatedness across Libraries}
\label{fig:UOL}
\end{subfigure}
\caption{Distributions of Usage Outdatedness across Projects and Libraries}
\end{figure*}

\textbf{Definition.} We first define the usage outdatedness of a library dependency $d$, denoted as $uso_d$, as the number of higher version releases of the library at the time of repository crawling. Hence, for each library $l \in \tilde{\mathcal{L}}_{us}$, we crawled the version number and release date of $l$'s all library version releases from library repositories. We had \todo{288,312} library version releases, denoted as $\mathcal{R}_{us}$. Each library version release $r \in \mathcal{R}_{us}$ is denoted as a 2-tuple $\langle v,$ $date \rangle$, where $v$ denotes a library version, and $date$ denotes $v$'s release date. Using $\mathcal{D}_{us}$ and $\mathcal{R}_{us}$, we compute $uso_d$ by Eq. \ref{eq:uso1}.

\begin{small}
\begin{equation}
\begin{aligned}
\forall d \in \mathcal{D}_{us}, uso_d = &|\{r | r \in \mathcal{R}_{us} \wedge r.v.l = d.v.l \wedge \\ & r.v.ver > d.v.ver \wedge r.date < d.p.date\}|
\end{aligned}\label{eq:uso1}
\end{equation}
\end{small}

Then, we define usage outdatedness from a project's and a library's perspective: $uso_p$, the average usage outdatedness of the library dependencies in a project $p$, and $uso_l$, the average usage outdatedness of the library dependencies on a library $l$. Using $\mathcal{P}$, $\tilde{\mathcal{L}}_{us}$, $\mathcal{D}_{us}$ and $uso_d$, we compute $uso_p$ and $uso_l$ by Eq. \ref{eq:uso2} where $\mathcal{D}_p = \{d | d \in \mathcal{D}_{us} \wedge d.p = p\}$, $\mathcal{D}_l = \{d | d \in \mathcal{D}_{us} \wedge d.v.l = l\}$.

\begin{small}
\begin{equation}
\begin{aligned}
\forall p \in \mathcal{P}, uso_p =& avg_{d \in \mathcal{D}_p} uso_d \\
\forall l \in \tilde{\mathcal{L}}_{us}, uso_l =& avg_{d \in \mathcal{D}_l} uso_d
\end{aligned}\label{eq:uso2}
\end{equation}
\end{small}


\textbf{Basic Findings.} Using $uso_p$ and $uso_l$, we show distributions of usage outdatedness across projects and libraries in Fig. \ref{fig:UOP} and \ref{fig:UOL}. On the one hand, only 28 (3.5\%) projects use the latest library versions. 83 (10.3\%) projects adopt libraries that are on average at most two versions away from the latest. 306 (38.0\%), 118 (14.6\%) and 19 (2.4\%) projects use libraries that are on average more than 10, 20 and 50 versions away from the latest, respectively. Notice that 33 projects are not included in Fig. \ref{fig:UOP} because 28 projects do not adopt libraries, and the jar files of the library versions in 5 projects are all no longer available. On the other hand, in all the projects that use them, 3,269 (45.2\%) libraries are already the latest. 1,419 (19.6\%) libraries are on average at most two versions away from the latest. 1,025 (14.2\%) and 134 (1.9\%) libraries are on average more than 10 and 50 versions away from the latest, respectively. These findings indicate that it is quite common to adopt outdated libraries and relatively often the distance to the latest version is considerably large. 


\textbf{In-Depth Findings.} Given the results of method-level usage intensity in Sec. \ref{sec:usageIntensityM}, it is reasonable for project developers to intentionally skip some library versions because the used part of library APIs (which is mostly the minority of all library APIs) is not changed and works as intended. As revealed by a recent survey \cite{Derr2017KMU}, this is actually the most common reason for adopting outdated libraries; and other main reasons include API incompatibilities, integration effort and unawareness of new versions. On the other hand, developers do update library versions, as 13.7\% projects adopt libraries close to the latest version. However, it is not clear how developers update library versions, which motivates our follow-up library update analysis in Sec. \ref{sec:update}.

We follow the same procedure as in Sec. \ref{sec:usageIntensityL} to analyze the categories of 50 libraries directly taken from the tail of Fig. \ref{fig:UOL} and 344 libraries statistically sampled from the first bar in Fig. \ref{fig:UOL}. Noticeably, almost half of the 50 libraries are cloud computing libraries, which could need huge integration effort to switch to new versions; and 130 of the 344 libraries are Eclipse plugin libraries, which have to be updated accordingly if developers want new features from and switch to new Eclipse versions.
 

\begin{figure*}[!t]
\centering
\begin{subfigure}[b]{0.45\textwidth}
\centering
\includegraphics[width=0.99\textwidth]{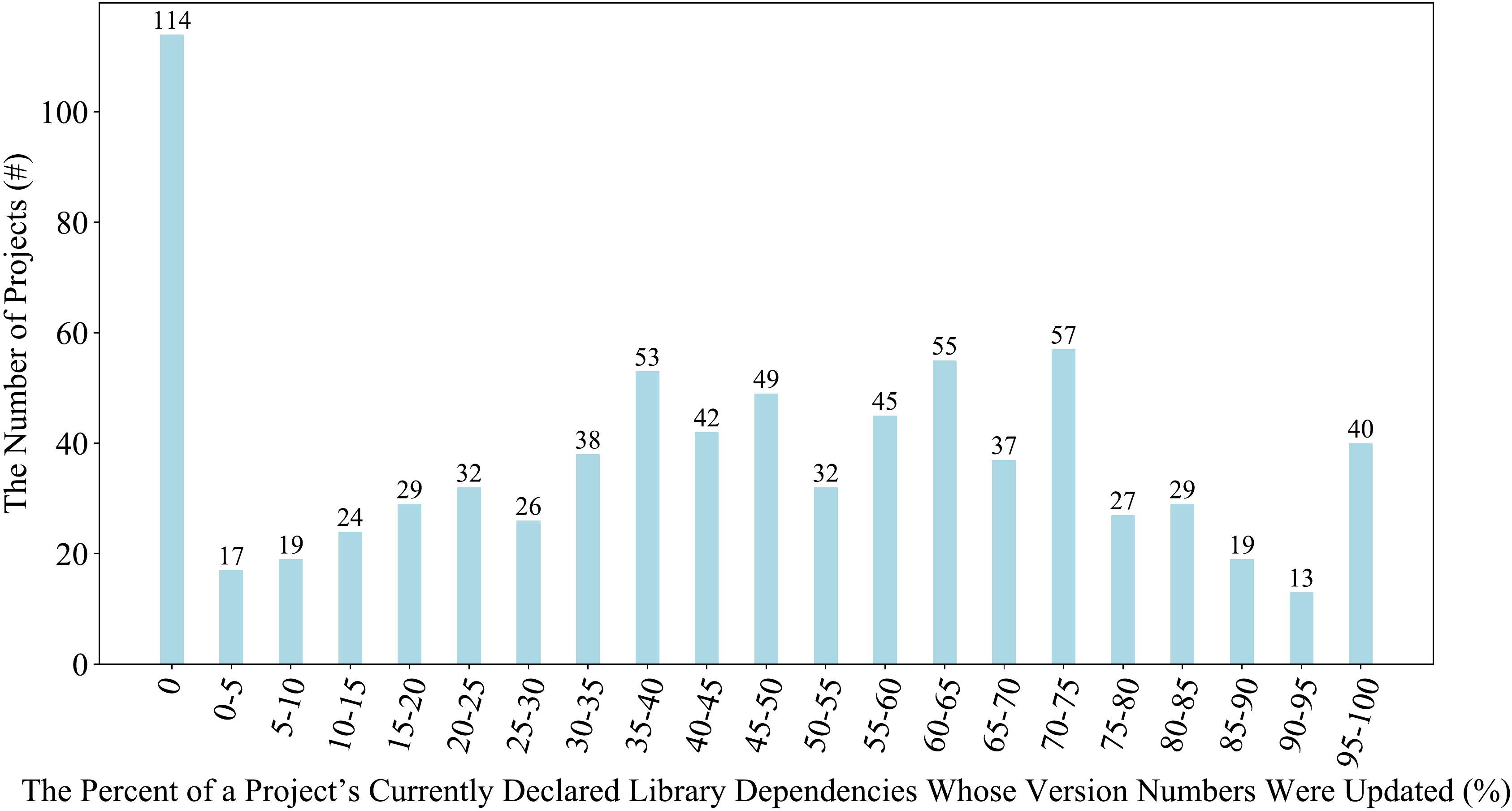}
\caption{Update Intensity across Projects}
\label{fig:intensityPro}
\end{subfigure}~~~
\begin{subfigure}[b]{0.45\textwidth}
\centering
\includegraphics[width=0.99\textwidth]{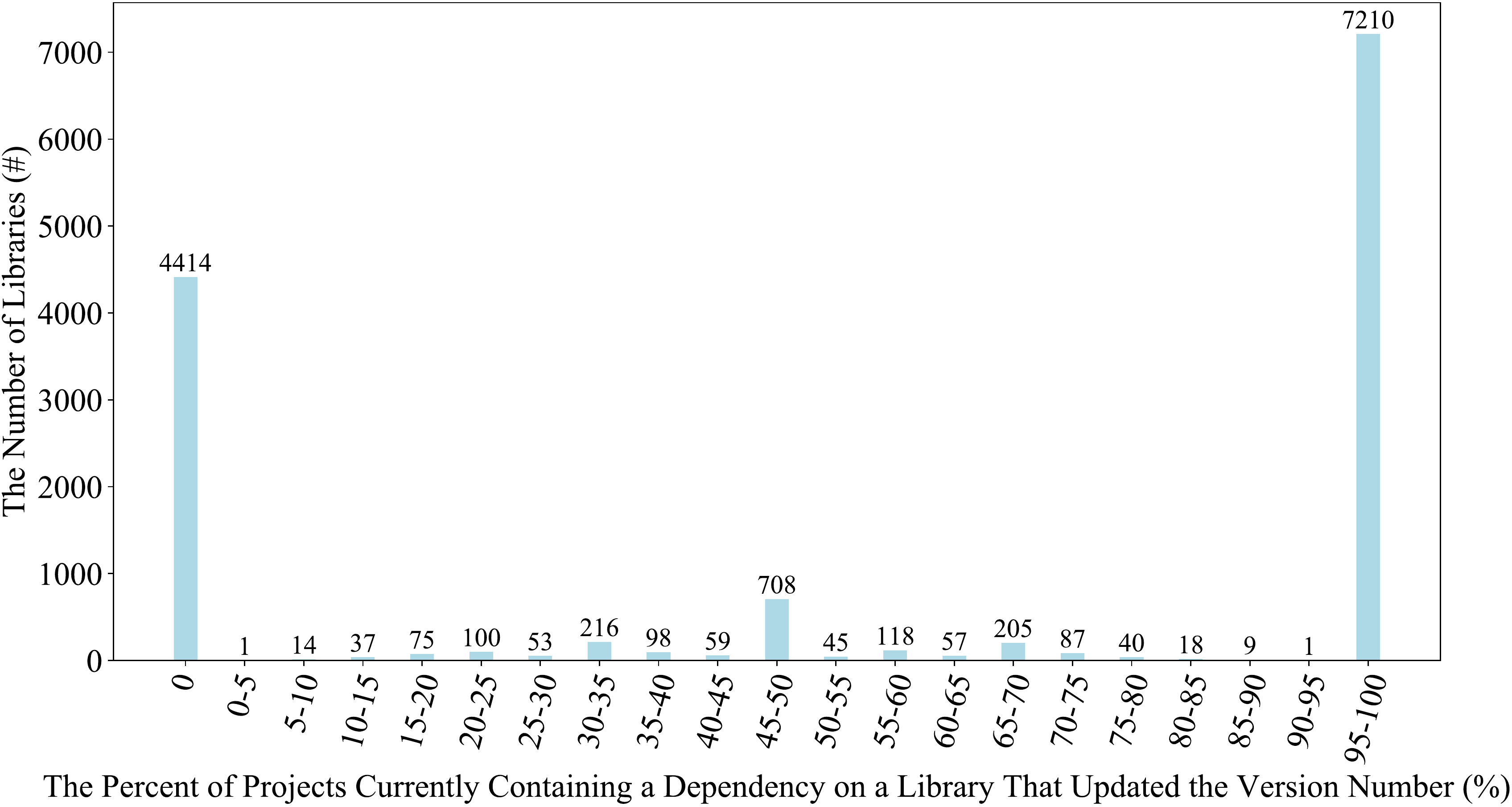}
\caption{Update Intensity across Libraries}
\label{fig:intensityLib}
\end{subfigure}
\caption{Distributions of Update Intensity across Projects and Libraries}
\end{figure*}

\section{Library Update Analysis}\label{sec:update}

To investigate library updates, we develop \textit{up-extractor} to extract library version updates from a project's commits. It scans a project $p$'s commits to locate every commit $com$ that changes $p$'s configuration file(s). Then, it uses \textit{lib-extractor} (see Sec. \ref{sec:usage}) to $com$ and $com$'s previous commit so as to respectively extract the library dependencies before and after $com$. Finally, from the two sets of library dependencies, it identifies library version updates by searching the library dependencies whose version number is changed. Each library version update $u$ is denoted as a 6-tuple $\langle p, f, com, l, ver_1, ver_2 \rangle$, where $p$, $f$, $com$ and $l$ respectively denote the project, configuration file, commit and library where $u$ occurs, and $ver_1$ and $ver_2$ respectively denote the version number before and after the update.

We used \textit{up-extractor} to the commits of each project $p \in \mathcal{P}$, and extracted \todo{5,217,348} library version updates, denoted as $\mathcal{U}$.

\subsection{Update Intensity}\label{sec:updateIntensity}

\textbf{Definition.} We define update intensity from a project's and a library's perspective: $upi_p$, the percent of a project $p$'s currently declared library dependencies whose version numbers were updated in $p$'s commits, and $upi_l$, the percent of projects currently containing a dependency on a library $l$ that updated $l$'s version number in their commits. Based on $\mathcal{P}$, $\mathcal{L}_{us}$, $\mathcal{D}_{us}$ and $\mathcal{U}$, we compute $upi_p$ and $upi_l$ by Eq \ref{eq:up}, where $\mathcal{D}_p = \{d | d \in \mathcal{D}_{us} \wedge d.p = p\}$ and $\mathcal{P}_l = \{d.p | d \in \mathcal{D}_{us} \wedge d.v.l = l\}$.

\begin{small}
\begin{equation}
\begin{aligned}
\forall p \in \mathcal{P}, upi_p =& |\{d | d \in \mathcal{D}_P \wedge (\exists u \in \mathcal{U}, u.p = d.p \\ &\wedge u.f = d.f \wedge u.l = d.v.l)\}| / |\mathcal{D}_p| \\
\forall l \in \mathcal{L}_{us}, upi_l = & |\{p | p \in \mathcal{P}_l \wedge (\exists u \in \mathcal{U}, u.p = p \\ & \wedge u.l = l)\}| / |\mathcal{P}_l|
\end{aligned}\label{eq:up}
\end{equation}
\end{small}


\textbf{Basic Findings.} Using $upi_p$ and $upi_l$, we show distributions of update intensity across projects and libraries in Fig. \ref{fig:intensityPro} and \ref{fig:intensityLib}. On the one hand, 114 (14.1\%) projects did not update any currently-declared library dependency, where 90 projects never updated any library dependency, and 24 projects updated library dependencies that were removed. 89 (11.0\%) and 329 (40.8\%) projects respectively updated at most 20\% and 50\% of their currently-declared library dependencies. 354 (43.9\%) and 101 (12.5\%) projects respectively updated more than 50\% and 80\% of their currently-declared library dependencies. Notice that 9 projects do not declare any library dependency and are not included in Fig. \ref{fig:intensityPro}. On the other hand, 4,414 (32.5\%) libraries were never updated in all the projects that depend on them. At the other extreme, 7,210 (53.2\%) libraries were updated in more than 95\% of the projects that use them. Such two extremes are mainly caused by the fact that 90.3\% of these libraries are only used by one project (as also evidenced in Fig. \ref{fig:LUIL}). If excluding all 10,499 libraries only used by one project, we find that of the remaining 3,066 libraries, 2,004 (65.4\%) libraries were not updated in more than half of the projects that adopt them. These findings show that developers do update library dependencies, but still i) 54.9\% projects leave more than half of their library dependencies never updated, and ii) one-third of libraries were not updated in more than half of the projects that use them.

\begin{figure}[!t]
\centering
\begin{subfigure}[b]{0.155\textwidth}
\centering
\includegraphics[width=0.99\textwidth]{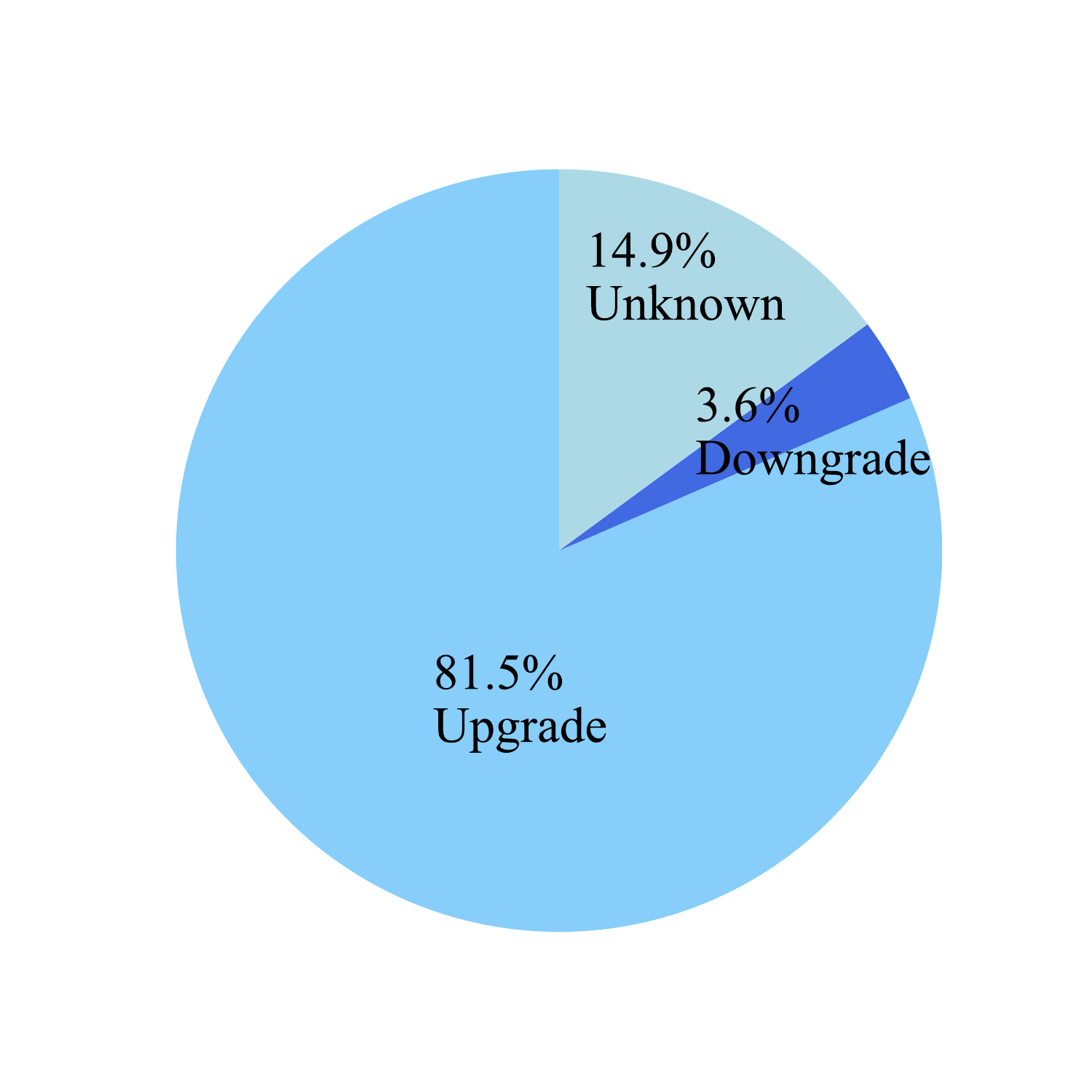}
\caption{Updates}
\label{fig:updatesDis}
\end{subfigure}%
\begin{subfigure}[b]{0.155\textwidth}
\centering
\includegraphics[width=0.99\textwidth]{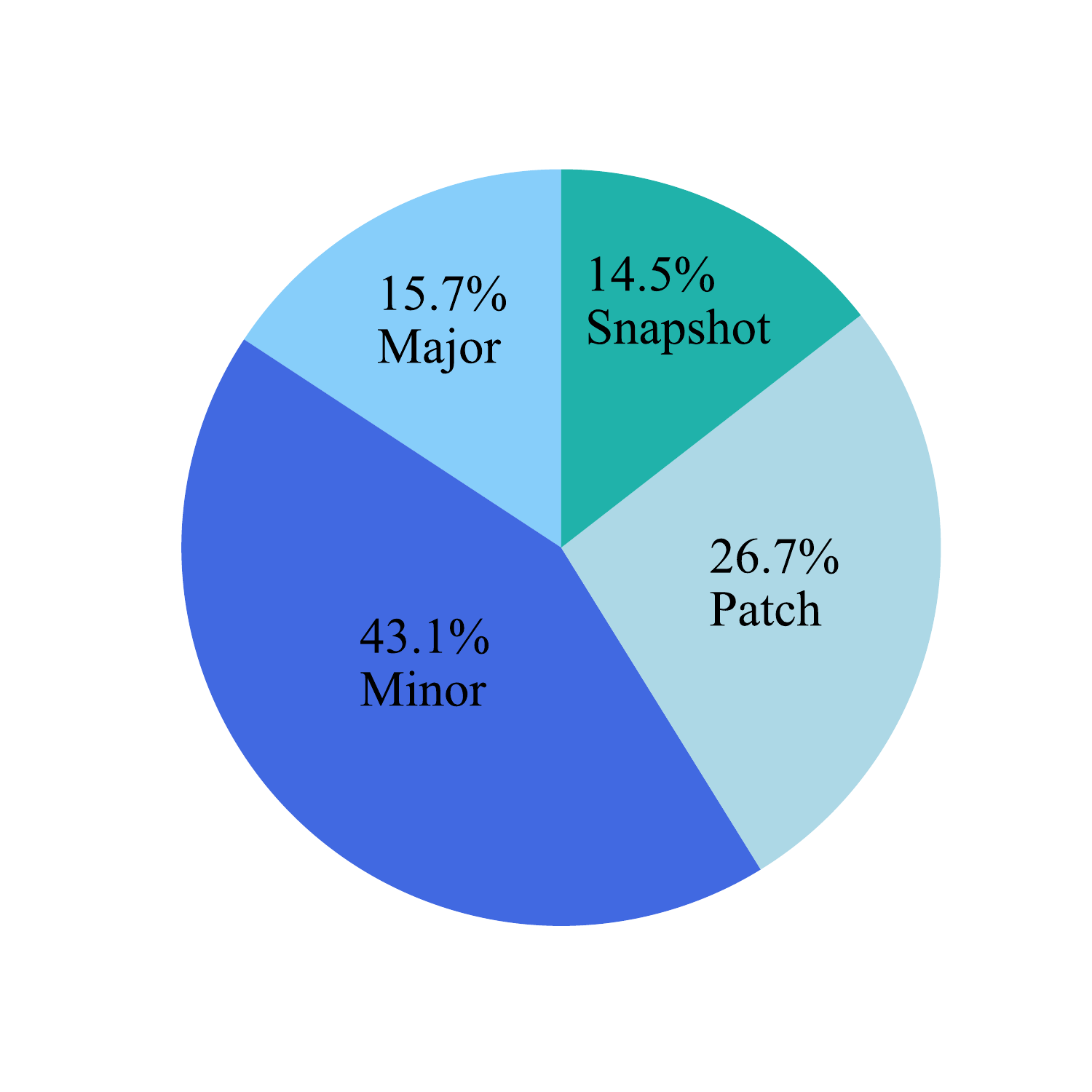}
\caption{Upgrades}
\label{fig:upgrade}
\end{subfigure}%
\begin{subfigure}[b]{0.155\textwidth}
\centering
\includegraphics[width=0.99\textwidth]{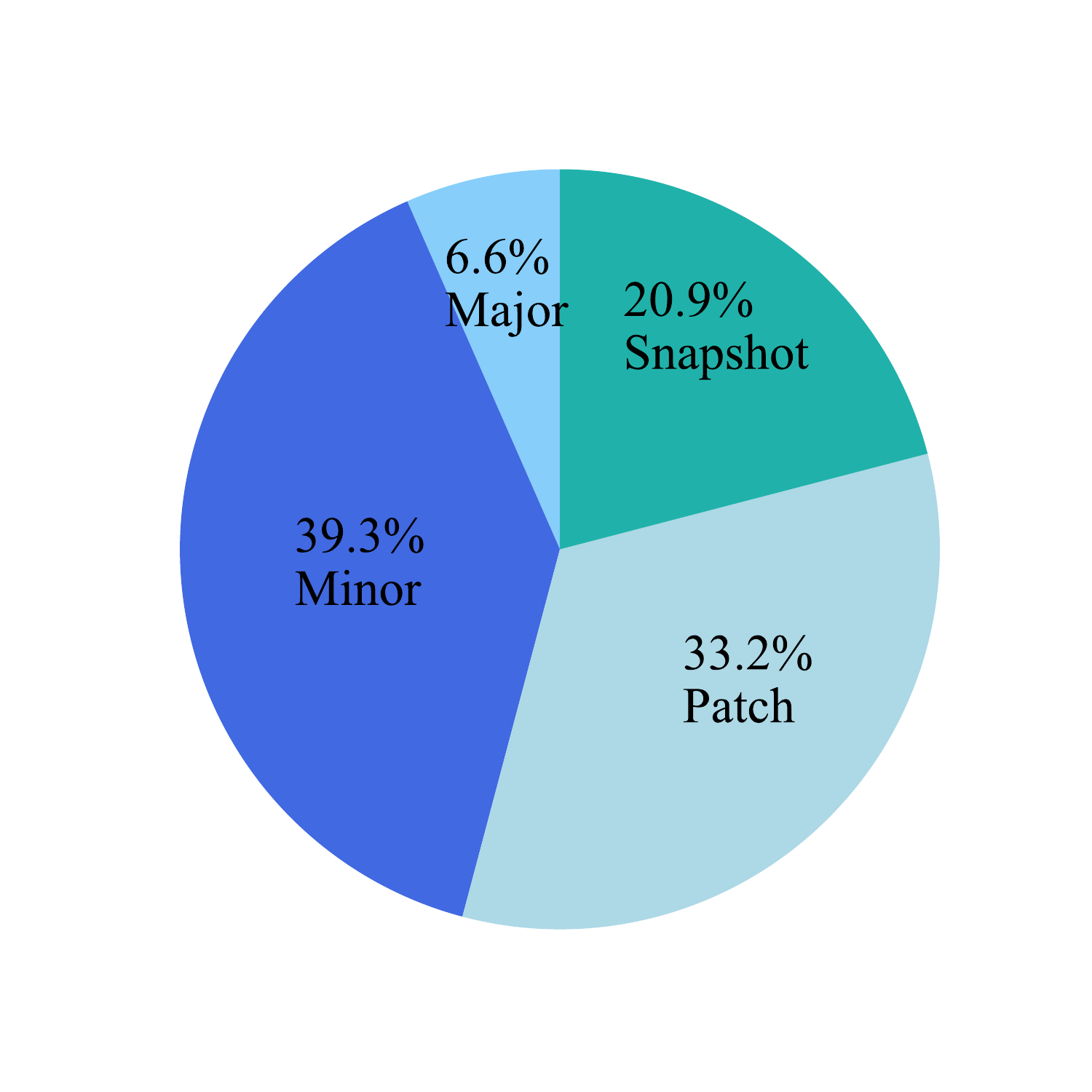}
\caption{Downgrades}
\label{fig:downgrade}
\end{subfigure}
\caption{Distributions of Updates, Upgrades and Downgrades}
\end{figure}

\textbf{In-Depth Findings.} Apart from the perspective of a project and a library, we further analyze the changes of version numbers in library version updates. Defined by semantic versioning \cite{preston2013semantic}, version numbers must take the form of $X.Y.Z$, where $X$, $Y$ and $Z$ is the major, minor and patch version. Bug fixes not affecting APIs increment $Z$, backwards compatible API changes or additions increment $Y$, while backwards incompatible API changes increment $X$. Generally, developers need no integration effort if updating to a patch or minor version, but need some integration effort if updating to a major version. We identified 5,117,870 (98.1\%) library version updates from $\mathcal{U}$ whose version numbers start with $X.Y$ or $X.Y.Z$, denoted as $\hat{\mathcal{U}}$.

On the one hand, we explore whether developers upgrade or downgrade a library version. As shown in Fig. \ref{fig:updatesDis}, most updates are upgrades; and a very small part (3.6\%) of updates are downgrades due to incompatible APIs. 14.9\% of them contain diverse suffixes in version numbers, and hence are unknown due to incomparable version numbers. On the other hand, we investigate whether developers update major, minor, patch, or snapshot versions, and report the results in Fig. \ref{fig:upgrade} and \ref{fig:downgrade}. First, 14.5\% of upgrades replace snapshot versions with stable versions due to the unstable nature of snapshot versions; and 20.9\% of downgrades switch back to snapshot versions because of heavy dependency on unstable APIs. Second, 79.8\% upgrades are minor or patch as they are supposed to be API compatible; and 72.5\% downgrades are also minor or patch because of violations of semantic versioning. Third, major upgrades or downgrades are less common as incompatible APIs can be introduced in major versions. These findings indicate that snapshot versions need to be better managed, semantic versioning should be followed but are not strictly followed (which is also evidenced in \cite{raemaekers2014semantic}), and major versions deserve a mechanism to be kept updated.

\subsection{Update Delay}\label{sec:updateDelay}

\begin{figure*}[!t]
\centering
\begin{subfigure}[b]{0.45\textwidth}
\centering
\includegraphics[width=0.99\textwidth]{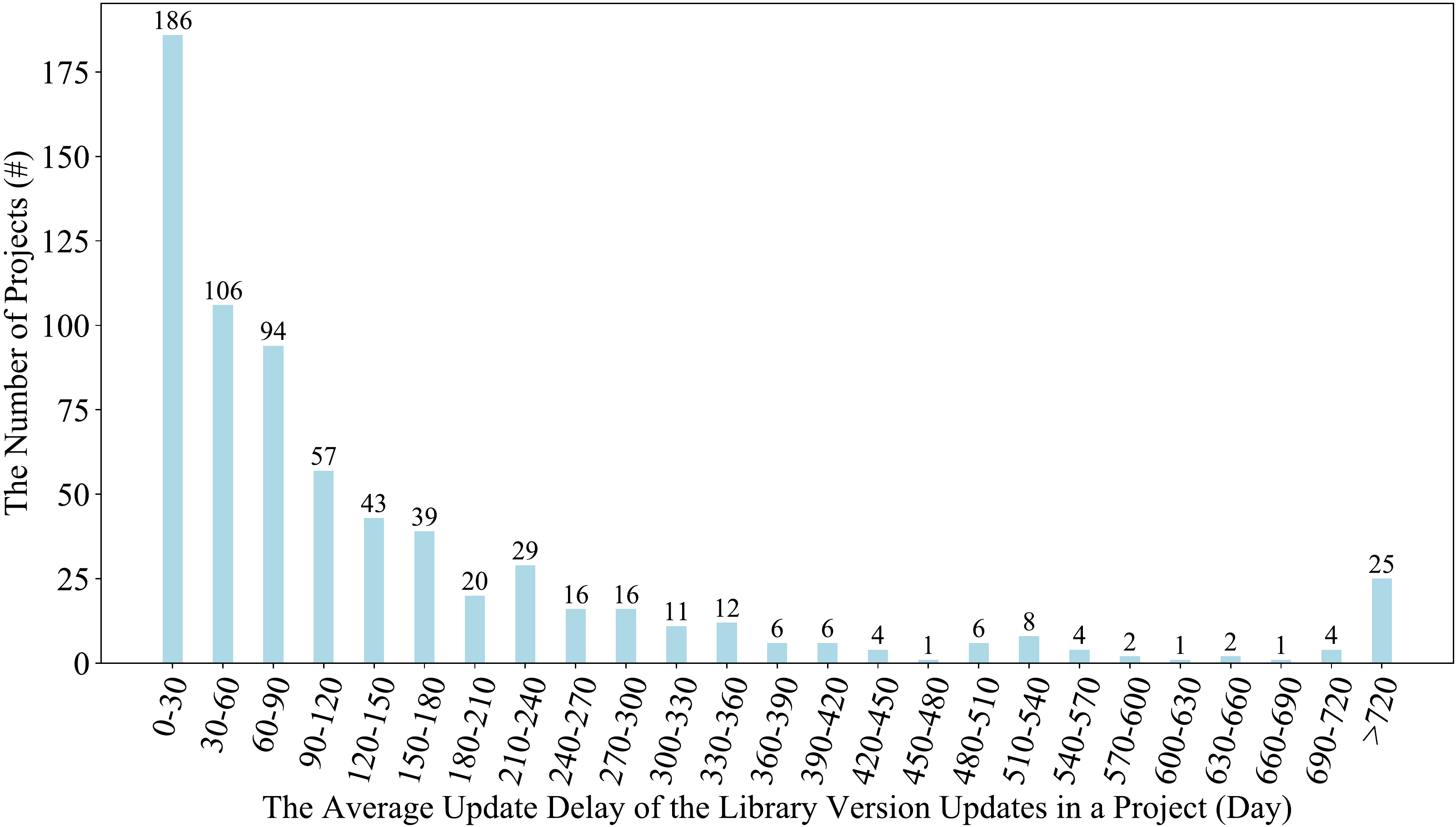}
\caption{Update Delay across Projects}
\label{fig:delayPro}
\end{subfigure}~~~
\begin{subfigure}[b]{0.45\textwidth}
\centering
\includegraphics[width=0.99\textwidth]{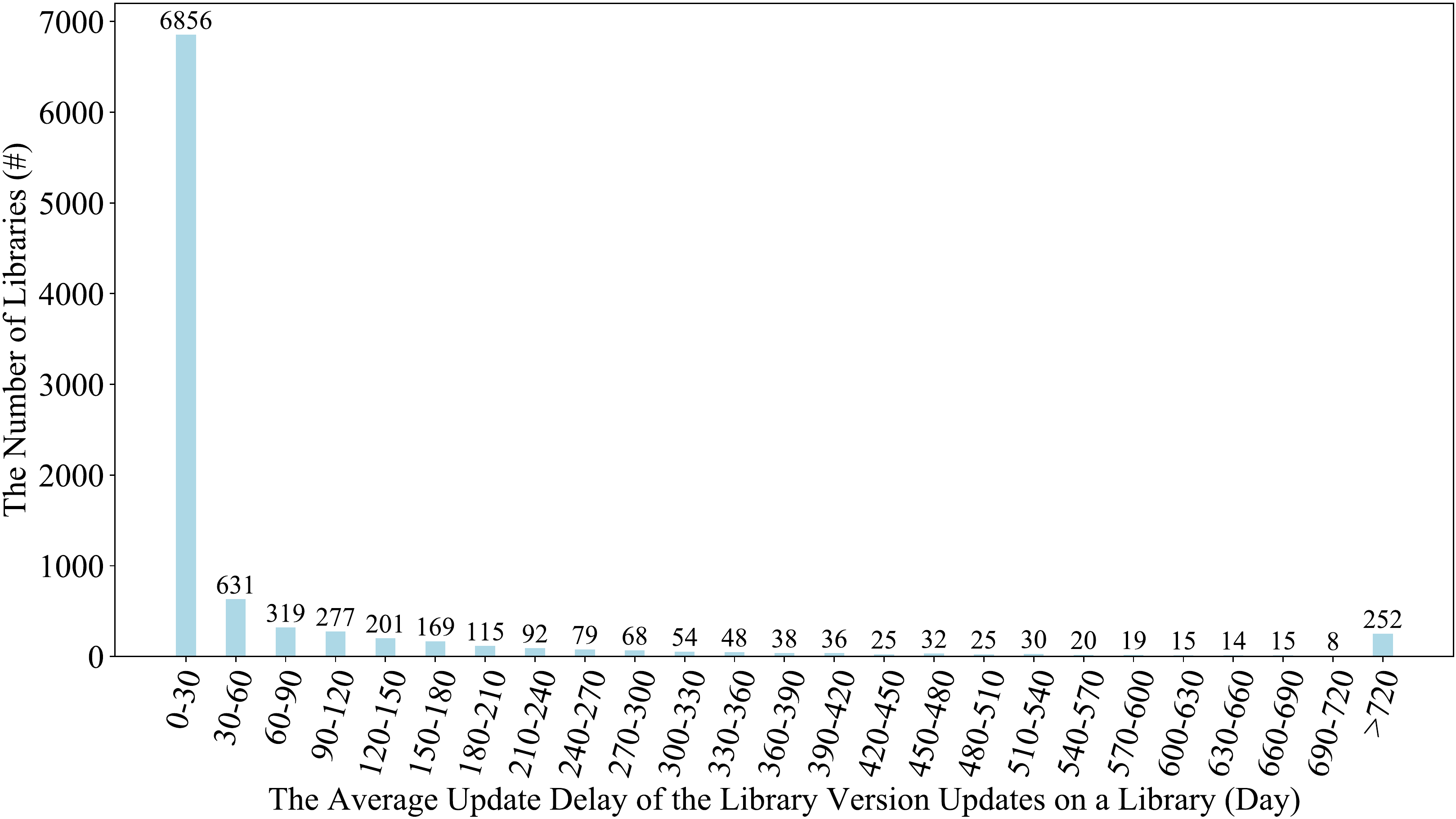}
\caption{Update Delay across Libraries}
\label{fig:delayLib}
\end{subfigure}
\caption{Distributions of Update Delay across Projects and Libraries}
\end{figure*}

\textbf{Definition.} We first define the update delay of a library version update $u$, denoted as $upd_u$, as the delay between the commit date of $u$ and the release date of the library version after $u$. Hence, for each library version update $u \in \mathcal{U}$, we crawled the release data of $\langle u.l, u.ver_2 \rangle$ from library repositories. We successfully crawled for \todo{1,507,196} (28.9\%) library version updates, denoted as $\tilde{\mathcal{U}}$, resulting in 155,969 library version releases, denoted as $\mathcal{R}_{up}$. From $\mathcal{R}_{up}$, we had \todo{9,438} libraries, denoted as $\mathcal{L}_{up}$ (i.e., $\mathcal{L}_{up} = \{r.v.l | r \in \mathcal{R}_{up}\}$). Of the library version updates we failed to crawl, 87.8\% are caused by snapshot versions that are no longer available. Using $\tilde{\mathcal{U}}$ and $\mathcal{R}_{up}$, we compute $upd_u$ by Eq. \ref{eq:upd}, where  $r \in \mathcal{R}_{up} \wedge r.v.l = u.l \wedge r.v.ver = u.ver_2$.

\begin{small}
\begin{equation}
\begin{aligned}
\forall u \in \tilde{\mathcal{U}}, upd_u = u.com.date-r.date
\end{aligned}\label{eq:upd}
\end{equation}
\end{small}

Then, we define update delay from the perspective of a project and a library: $upd_p$, the average update delay of the library version updates in a project $p$, and $upd_l$, the average update delay of the library version updates on a library $l$. Using $\mathcal{P}$, $\mathcal{L}_{up}$, $\tilde{\mathcal{U}}$ and $upd_u$, we compute $upd_p$ and $upd_l$ by Eq. \ref{eq:updd}, where $\mathcal{U}_p = \{u | u \in \tilde{\mathcal{U}} \wedge u.p$ $= p\}$ and $\mathcal{U}_l = \{u | u \in \tilde{\mathcal{U}} \wedge u. l = l\}$.

\begin{small}
\begin{equation}
\begin{aligned}
\forall p \in \mathcal{P}_{up}, upd_p = &avg_{u \in \mathcal{U}_p} upd_u \\
\forall l \in \mathcal{L}_{up}^\prime, upd_l = &avg_{u \in \mathcal{U}_l} upd_u
\end{aligned}\label{eq:updd}
\end{equation}
\end{small}

\textbf{Basic Findings.} Using $upd_p$ and $upd_l$, we show distributions of update delay across projects and libraries in Fig. \ref{fig:delayPro} and \ref{fig:delayLib}. On the one hand, 186 (23.1\%) projects updated their library dependency at a lag of at most 30 days. 407 (50.5\%), 256 (31.8\%) and 174 (21.6\%) projects had an update delay of more than 60, 120 and 180 days. Notice that 107 (13.3\%) projects are not included in Fig. \ref{fig:delayPro} as we failed to compute the update delay (90 projects never updated any library dependency; and 17 projects updated library dependencies but we failed to crawl the release date). On the other hand, 6,856 (72.6\%) libraries were updated at a lag of at most 30 days. 1,951 (20.7\%), 1,355 (14.4\%) and 985 (10.4\%) libraries had an update delay of more than 60, 120 and 180 days. These findings indicate that i) project developers mostly have slow reaction to new library version releases, and ii) libraries are not updated in a timely way.

\textbf{In-Depth Findings.} We further analyze whether the update delay of a library version update $u$ matters by checking whether the library APIs of $\langle u.l, u.ver_1 \rangle$ called in project $u.p$ are not changed in $\langle u.l, u.ver_2 \rangle$. Intuitively, if all called library APIs are not changed, the update delay of $u$ will not have any severe influence; e.g., developers may update a library for new features. Here, we regard a library API as not changed if the code of the API as well as the code of the methods in its call graph are not changed. Because of the heavy computation involved in this analysis, we only target \todo{14,572} library version updates in $\tilde{\mathcal{U}}$ that occur in recent three months and whose library versions can be crawled. Surprisingly, the update delay of \todo{78.9\%} library version updates does not matter. It is reasonable as developers are often unwilling to update as changed APIs may break \cite{Derr2017KMU}. However, it is still not clear what the update delay of the mattering updates means, which motivates our library risk analysis in Sec. \ref{sec:risk}.


\section{Library Risk Analysis}\label{sec:risk}

As bug fixing is recognized as the most common reason for updating libraries \cite{Derr2017KMU}, we analyze library risks in terms of bugs. Hence, we develop \textit{bug-crawler} to crawl \textit{severe} bugs in library version releases from \textit{issue trackers}. We currently support the issue tracker Jira, and regard bugs whose priority is major, critical or blocker but not minor and trivial as severe bugs. In detail, it crawls the metadata of Jira issues of a library, and selects issues with type of bug, priority of major, critical or blocker, and status of closed and fixed. For each selected bug issues, it extracts the issue id, the priority, and the library version releases affected. 
We represent a bug in a library version release as a bug-release pair $g$, denoted as a 2-tuple $\langle b, r\rangle$, where $b$ denotes a bug, and $r$ denotes a library version release that $b$ affects. Each bug $b$ is denoted as a 2-tuple $\langle id, pri \rangle$, where $id$ denotes the issue id and $pri$ denotes the priority of $b$.

To make our library risk analysis in RQ3 feasible, we focused on the 50 most popular libraries derived from Sec. \ref{sec:usageIntensityL}. Of the 50 libraries, only 15 libraries, denoted as $\mathcal{L}_{ri}$, use Jira as the issue tracker. These 15 libraries have 722 library version releases (computed from $\mathcal{L}_{ri}$ and $\mathcal{R}_{us}$). We applied \textit{bug-crawler} to each library $l \in \mathcal{L}_{ri}$, and crawled \todo{1,170} bugs (i.e., 1,074 major bugs, 66 critical bugs and 30 blocker bugs). Finally, we had \todo{1,432} bug-release pairs, denoted as $\mathcal{G}$. From $\mathcal{G}$, we had \todo{228} buggy library version releases, denoted as $\mathcal{R}_{ri}$ (i.e., $\mathcal{R}_{ri} = \{g.r | g \in \mathcal{G}\}$).

\textbf{Definition.} We define the potential risk of a library version release $r$, denoted as $rib_r$, as the number of severe bugs in $r$. Using $\mathcal{R}_{ri}$ and $\mathcal{G}$, we compute $rib_r$ by Eq. \ref{eq:rib}. 

\begin{small}
\begin{equation}
\begin{aligned}
\forall r \in \mathcal{R}_{ri}, rib_r = |\{g.b | g \in \mathcal{G} \wedge g.r = r\}|
\end{aligned}\label{eq:rib}
\end{equation}
\end{small}


\begin{figure}[!t]
\centering
\includegraphics[width=0.42\textwidth]{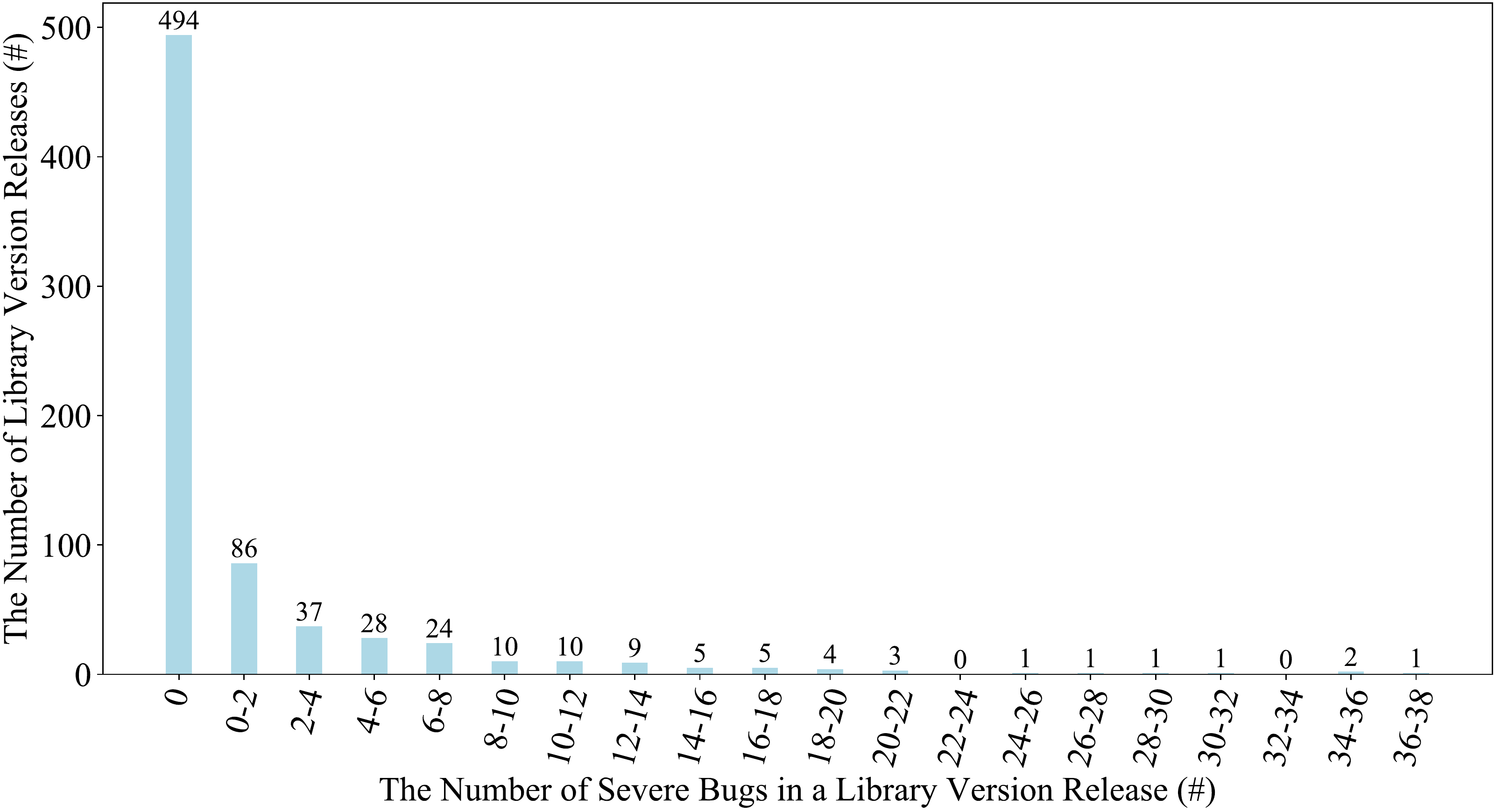}
\caption{Potential Risk across Library Version Releases}
\label{fig:risk}
\end{figure}

\textbf{Basic Findings.} Using $rib_r$, we report the distribution of potential risk across library version releases in Fig. \ref{fig:risk}. 228 (31.6\%) library version releases of the 15 libraries have severe bugs. 105 (14.5\%) and 43 (6.0\%) library version releases have more than four and ten severe bugs, respectively. Though only considering severe bugs, these findings indicate the relatively common existence of bugs in popular libraries and thus the potential risk of affecting client projects if developers are unaware of the bugs in libraries or delay library updates (as indicated in Sec. \ref{sec:updateDelay}).


\textbf{In-Depth Findings.} To characterize developers' response to buggy library versions, we submit issues by reporting the buggy library versions projects adopted and the set of severe bugs in them. Hence, we first identify the buggy library versions that a project $p \in \mathcal{P}$ uses, denoted as $\mathcal{V}_{ri}^p$ (i.e., $\mathcal{V}_{ri}^p =\{d.v | d \in \mathcal{D}_{us} \wedge$ $d.p = p \wedge (\exists r \in \mathcal{R}_{ri}, r.v = d.v)\}$). Then, for each library version $v \in \mathcal{V}_{ri}^p$, we find the set of severe bugs in $v$, denoted as $\mathcal{B}_v$ (i.e., $\mathcal{B}_v = \{g.b | g \in \mathcal{G} \wedge g.r.v = v\}$).

Totally, we identified \todo{433} projects that adopted buggy library versions. Of these projects, \todo{8} already updated the buggy library versions at the time of our issue reporting, \todo{73} disable issues and also do not use other issue trackers, and \todo{6} are read-only. As a result, we submitted an issue for \todo{346} projects, and received responses from \todo{78} (\todo{22.5}\%) issues. Two of the authors individually analyzed the issue responses to categorize the reaction types to our issues, and then conducted a group discussion to summarize the reaction types. After reaching a consensus, they reviewed and relabeled the types of reactions in issue responses together.

After this manual analysis, we summarized \todo{7} reactions types. First, developers directly closed the issue without any comment in 19 (24.4\%) issues. Second, developers closed the issue and regarded it as spam in 9 (11.5\%) issues. These two types indicate that developers regard our issues as not helpful, or do not care about buggy library versions. Third, developers confirmed the issue and will fix it when new library versions are released since they already use the latest library versions, which accounts for 5 (6.4\%) issues. Fourth, developers confirmed the issue and will fix it later since it takes some integration effort to update library versions, which happens in 9 (11.5\%) issues. Fifth, developers directly fixed the issue by updating library versions in 8 (10.3\%) issues. These three types show that developers regard our issues as helpful, and are willing to update library versions. Sixth, developers will not fix the issue as the projects heavily depend on the buggy library versions and it will take huge effort to update them, which appears in 3 (3.8\%) issues. This type reflects that developers may consider our issues as useful, but decide to take the risk due to the huge fix effort. Seventh, developers suggested us to improve our issue quality, as the buggy library versions are only used in test code in 11 (14.1\%) issues, the bugs are not in projects' execution paths in 8 10.3\%) issues, the bugs are not severe enough for them to fix and they are more interested in security bugs in 2 (2.6\%) issues, and the buggy library versions are declared as optional in the projects and will be re-declared when the projects are used by other projects in 2 (2.5\%) issues. This type demonstrate that developers need more convincing and fine-grained information to decide whether to update buggy library versions. For the remaining 2 issues, one thanks for our issue but does nothing, and the other searches GitHub to locate our tool. These findings classify developers' reactions to buggy library versions, and provide practical implications (see Sec. \ref{sec:implication}) on alerting systems that are acceptable by project developers.



\section{Implications, Application and Threats}\label{sec:implication}

In this section, we elaborate the practical implications of our findings, demonstrate the usefulness of our findings by a prototype application, and discuss the threats to our study.

\subsection{Implications to Researchers and Developers}

\textbf{Library Debloat.} As a very small portion of library APIs are widely called across projects, many unused library features are still kept in software systems, which could cause software bloat, especially for embedded systems. Software bloat could hurt performance \cite{Xu2010SBA} or broaden attack surface (e.g., code reuse attack) \cite{Wang2019reuse}. Therefore, the small usage intensity of library APIs opens an opportunity to eliminate unused features in libraries (i.e., library debloat) in their usage context to avoid software bloat. Recently, some debloat techniques \cite{Quach2018, sharif2018trimmer} have been designed for C/C++, while there is still no practical technique for Java.


\textbf{Multiple Version Harmonization.} Multiple versions of the same library are commonly used in one project. It may increase the burden of project developers as multiple analyses are needed to learn the differences of these multiple versions, or even cause dependency conflicts \cite{Wang2018DCM, patra2018conflictjs}. Therefore, techniques are needed to automatically identify multiple versions, analyze their differences in client usage context, and refactor client code and configuration files to harmonize to one single version. While it might be a tedious task to unify multiple versions, we believe it brings sustainable benefits in library maintenance in the long run. 

\textbf{Snapshot Version Management.} Snapshot library versions are widely used. They are in fact unstable or unfinished versions that are still under heavy development, which increases both the maintenance cost and the risk of incompatible APIs. Therefore, it is a double-edged sword; i.e., it could keep projects using the latest library versions, but might cause projects depend on incompatible APIs. In that sense, it is worthwhile for researchers deeply investigating the benefits and costs of snapshot versions to shed light on the management of snapshot versions.

\textbf{Smart Alerting and Automated Updating.} Given the commonness of buggy, outdated libraries in projects, it is urgent to propose techniques to alert and update buggy, outdated libraries. Our analysis of developers' responses to buggy libraries implies developers' requirements of such an alerting system. On the one hand, alerts should be raised only when bugs in library versions are in execution paths of projects. Otherwise, buggy library versions are safe. On the other hand, multiple fine-grained information should be provided to assist developers to make confident decisions in updating buggy library versions. Specifically, alerts should indicate whether the production or test code is affected, whether the bugs are security or non-security bugs, and the statistics about  the library API calls affected by the bugs, such that developers can assess risks. Alerts should also report the statistics about the calls to library APIs  changed in the new library version, such that developers can assess the efforts to complete the update. Unfortunately, none of the alerting systems from academics (e.g., \cite{cadariu2015tracking, PontaPS2018beyond}) and industries (e.g., \cite{blackduck, sourceclear, snyk, greenkeeper}) can provide all these fine-grained information, and they only focus on security bugs.

\textbf{Automated Library Bug Triggering.} Bugs commonly exist in library versions. However, developers are unwilling to update libraries. Therefore, techniques are needed to automatically determine whether a bug in a library version can be triggered in the client code of a project. Such techniques can give developers motivated drivers to update libraries and improve project quality.

\textbf{Usage-Driven Library Evolution.} Our method-level library usage analysis also presents an opportunity for library developers to conduct usage-driven library evolution (e.g., assigning high fix priority to bugs in widely-used APIs, carefully evolving widely-used APIs via considering their change impacts on client projects, and assessing whether new APIs are adopted). Further, our findings can raise the attention of developers on problems in library maintenance (e.g., multiple, snapshot or buggy versions).

\textbf{Dataset and Visualization.} Our study produces a lot of data: declared library dependencies, library APIs, library API calls, library version releases, library version updates as well as bugs in library versions. We released them at \url{https://3rdpartylibs.github.io} together with our analysis tools to ease the reproduction and foster valuable applications. One potential application is to use or develop visualization techniques \cite{Kula2014VES, kula2018generalized} to visualize our data to assist project and library developers to make decisions in evolving, adopting and updating libraries.


\subsection{Application for Usefulness Demonstration}

According to the implication on smart alerting, we develop a prototype of a bug-driven alerting system for buggy libraries. It consists of two main components, {\it risk analysis} and {\it effort analysis}, and it have two databases: bug database and library database. The bug database contains the bugs in 15 popular libraries (see Sec. \ref{sec:risk}) as well as the corresponding buggy library methods (i.e., the methods changed in the patches that fix the bugs) in library versions; and the library database contains the jar files of all the released versions of the 15 popular libraries.

The risk analysis component decides whether a project could directly or indirectly call buggy library methods. It first extracts library API calls in the project, then constructs the call graphs of these called library APIs, and finally checks whether the library methods in each call graph contain buggy library methods in our bug database. If yes, we regard the called library API as buggy and affecting the project (i.e., the corresponding bug may in the execution path of the project). Hence, we can report the number of bugs that affect the project in each buggy library version (i.e., NB), the number of buggy library APIs called in the project (i.e., NA), and the number of calls to buggy library APIs in the project (i.e., NC). These three metrics provide developers with the risk and impact of buggy library versions.

\begin{table}[!t]
\scriptsize
\centering
\caption{Results of Applying Our Alerting System}\label{table:tool}
\begin{tabular}{|c||c||c|c|c||c||c|c|c|c|}
\hline
P & BL & NB & NA & NC & SL & NAD & NAC & NCD & NCC \\\hline
1 & 1 & 1(15) & 2(3) & 2(3) & 7 & 0 & 3 & 0 & 3 \\\hline
2 & 1 & 2(36) & 1(13) & 1(18) & 7 & 0 & 2 & 0 & 2 \\\hline
3 & 1 & 1(18) & 1(21) & 7(181) & 15 & 0 & 15 & 0 & 144 \\\hline
4 & 1 & 1(18) & 1(17) & 1(79) & 15 & 17 & 0 & 79 & 0 \\\hline
5 & 1 & 1(15) & 1(3) & 1(9) & 7 & 0 & 3 & 0 & 9 \\\hline
6 & 1  & 1(1) & 1(10) &1(13) & 9 & 0 & 1 & 0 & 1 \\\hline
7 & 1 & 1(15) & 3(3) & 9(9) & 7 & 0 & 3 & 0 & 9 \\\hline
8 & 1 & 1(11) & 1(8) & 2(24) & 1 & 0 & 3 & 0 & 6 \\\hline
9 & 1 & 1(15) & 1(3) & 1(7) & 7 & 1 & 1 & 1 & 5 \\\hline
\end{tabular}
\end{table}

The effort analysis component suggests the new library versions and their integration effort. For each of the higher library versions than the buggy library version, it first locates the called library APIs that are deleted/changed in the higher library version. Here, an API is changed if the body code of the API or the code of the library methods in its call graph is changed. Then, it checks whether the called library APIs that are not deleted can directly or indirectly call buggy library methods in the higher library version. If yes, we skip this higher library version because it still contains bugs affecting the project. If no, we can report the number of called library APIs deleted (i.e., NAD), the number of called library APIs changed (i.e., NAC), the number of calls to the deleted library APIs (i.e., NCD), and the number of calls to the changed library APIs (i.e., NCC). These metrics measure the integration effort on the suggested library version.

We have run our alerting system against the \todo{433} projects that use buggy libraries (see Sec. \ref{sec:risk}) to determine whether the buggy libraries affect the projects. We find that 424 projects are not affected by the buggy libraries and can be safe. For the 9 unsafe projects, we report the detailed results in Table \ref{table:tool}, where column \textit{P} lists the 9 projects, \textit{BL} reports the number of buggy libraries affecting the project, \textit{NB}, \textit{NA} and \textit{NC} are the reported metrics in risk analysis step (where the total number of bugs, called library APIs, and calls to the library APIs are listed in parentheses), \textit{SL} reports the number of suggested library versions, and the other columns report the metrics in effort analysis step for one of the suggested library versions (\todo{others are available at our website}). 

All the \todo{9} projects are affected by only one buggy library version. Although there are many bugs in the buggy library version, only one or two bugs affect at most three library APIs called by at most nine times in the project. For example, in the third project, 1 of the 21 called library APIs is affected by 1 bug, and is called by 7 times. Multiple higher library versions are suggested for developers to choose according to integration effort. As an example, 15 versions are suggested for  the third project. In one of them, 0 and 15 of the 21 called library APIs are respectively deleted and changed, affecting 0 and 144 library API calls. We submitted issues with such detailed reports, but have not received any reply (they also did not reply our issue in Sec.~\ref{sec:risk}).

We are enlarging our bug and library database, and collaborating with our industrial partner to include around 6K security bugs and integrate our prototype into their commercial tool. We are also enhancing the prototype in multiple dimensions, e.g, improving call graph precision and identifying API breakings.

\subsection{Threats to Validity}


\textbf{Indirect Library Dependency.} Our analyses are focused on direct library dependencies, i.e., libraries that are directly declared in the configuration file. Some libraries may depend on other libraries, i.e., indirect library dependencies. It can be expected that, if further considering indirect library dependencies, the dependency on libraries can be heavier, the potential risk in terms of bugs can be higher, and the problem of using multiple versions of the same library can be more severe \cite{Wang2018DCM}. We believe our findings are still representative if only considering direct library dependencies, and we will include indirect ones in future.

\textbf{Subject Representativity.} Our study involves various subjects: projects, jar files and release dates of libraries, and bugs in libraries. We choose active projects as library usage and update are software maintenance activities and inactive projects might contain less representative maintenance data and bias our findings. We fail to crawl jar files and release dates of some libraries, and hence they are excluded from some of our analyses, but we have tried our best and clarified the data for each analysis. We believe our data is still representative and meaningful due to the large size. We focus on 15 popular libraries that use Jira as the bug tracker. While the size is small, they are all widely used and worth the investigation. We are continuing library bug crawling to include more libraries and support more bug tracker.

\textbf{Library APIs.} We conservatively consider public methods and fields in public classes as library APIs. Hence, some public methods and fields that are not meant to be used by client projects are also treated as library APIs. Therefore, the real method-level usage intensity can be higher than reported. However, the only ground truth is in the documentations for library version releases, and is not always available.


\section{Related Work}\label{sec:related}

We review the most closely related work on library in five aspects: usage analysis, update analysis, risk analysis, evolution and adaptation, and recommendation and migration.

\subsection{Usage Analysis} 

Mileva et al. \cite{mileva2009mining} studied library version usage and analyze the usage trend and popularity of a library version and the times developers switched back from a library version. Similarly, Kula et al. \cite{kula2017exploratory} modeled the usage trend of a library version. 
Then, Mileva et al. \cite{Mileva2010} and Hora and Valente \cite{hora2015apiwave} investigated the usage trend and popularity of library API elements (classes and interfaces) via mining \textit{import} statements. These approaches analyzed coarse-grained library usage from a library's perspective. Instead, Lammel et al. \cite{lammel2011large} and De Roover et al. \cite{de2013multi} conducted library usage analysis at a fine-grained API method level from both a project's and a library's perspective, but did not distinguish library versions. Qiu et al. \cite{qiu2016understanding} also studied library API usage but only from the perspective of a library. In summary, the existing library usage analysis provides partial facets about library usage. To the best of our knowledge, this is the first work to \textit{holistically} analyze library usage at both a coarse-grained and fine-grained level from the perspective of a project and a library and provide practical implications to developers and researchers.

Bauer et al. \cite{bauer2012structured, bauer2012understanding} extracted the dependency of a project on library API methods. Similarly, Zaimi et al. \cite{zaimi2015empirical} computed the number of used library versions and library classes for a project. They analyzed library usage for one project, but did not aim at the mining of library usage knowledge across a corpus. Kula et al. \cite{kula2015trusting} analyzed the adoption of latest library versions when developers introduced libraries, and found that 82\% of projects adopted the latest version. Quantitatively, Cox et al. \cite{cox2015measuring} introduced three metrics to define the dependency freshness at the dependency and project level. In our study, we use one of the metrics to quantify usage outdatedness. Saied et al. \cite{saied2018improving} proposed an automatic approach to identify third-party library usage patterns (i.e., sets of libraries that are commonly used together). 

Apart from the Java ecosystem, studies have been conducted for the npm and Android ecosystems. Wittern et al. \cite{wittern2016look} investigated the popularity of npm packages and the adoption of semantic versioning in npm packages. Abdalkareem et al. \cite{abdalkareem2017developers} studied reasons and drawbacks of using trivial npm packages by a survey. In Android apps, library code is shipped into APK files, and thus library detection approaches \cite{Ma2016LFA,Li2017LSP,Backes2016RTL,zhang2018detecting} have been developed to improve clone detection \cite{chen2014achieving, wang2015wukong, linares2014revisiting}, library sandbox \cite{shekhar2012adsplit, seo2016flexdroid}, and malware detection \cite{li2016investigation}. Li et al. \cite{li2016investigation} also analyzed the popularity of mobile libraries. It is interesting to conduct fine-grained library usage in these ecosystems.

\subsection{Update Analysis} 

Bavota et al. \cite{bavota2013evolution, bavota2015apache} analyzed when and why developers updated inter-dependencies, and they found that a high number of bug fixes could encourage dependency updates, and API changes could discourage dependency updates. Fujibayashi et al. \cite{fujibayashi2017does} studied the relationship between library release cycle and library version updates. Kula et al. \cite{kula2018developers} analyzed the practice of library updates, and found that developers rarely updated libraries. They also conducted eight manual case studies to understand developer's responsiveness to new library version releases and security advisories, and found that developers were not likely to respond to security advisories mostly due to the unawareness of vulnerable libraries. Different from these studies, our study quantifies update intensity and update delay from the perspective of a project and a library.

Besides the Java ecosystem, library update analysis has been conducted for other ecosystems. Derr et al. \cite{Derr2017KMU} investigated why developers updated mobile libraries, studied the practice of semantic versioning, and conducted a library updatability analysis. Salza et al. \cite{Salza2018DUT} analyzed the mobile library categories that were more likely to be updated, and identified six update patterns. Lauinger et al. \cite{lauinger2017thou} and Zerouali et al. \cite{zerouali2018empirical} measured the time lag of an outdated npm package from its latest release, and Decan et al. \cite{Decan2018evolution} analyzed the evolution of this time lag. Decan et al. \cite{decan2017empirical} also compared problems and solutions of library updates  in three ecosystems, and found that the problems and solutions varied from one to another, and depended both on the policies and the technical aspects of each ecosystem.

\subsection{Risk Analysis} Decan et al. \cite{Decan2018ISV} studied the risk and impact of security bugs in npm packages. Similarly, we analyze the risk of bugs in Java libraries. We also classify developers' reactions to buggy libraries to give implications on smart alerting.

Cadariu et al. \cite{cadariu2015tracking} introduced an alerting system to report Java library dependencies having security bugs. Mirhosseini and Parnin \cite{mirhosseini2017can} studied the usage of pull requests and badges to notify outdated npm packages. Such alerting systems are very coarse-grained because they do not analyze whether bugs or code changes in library versions really affect a project. This was evidenced in a recent study, where Zapata et al. \cite{Zapata2018smoother} manually analyzed whether 60 projects called the npm packages' functions that were affected by security bugs, and found that 73.3\% of projects were actually safe from the security bugs. 

To mitigate this problem in previous alerting systems, several advances have been proposed to analyze whether bugs or code changes in libraries are truly in the execution path of a project. Hejderup et al. \cite{Hejderup2018SEC} constructed a versioned ecosystem-level call graph for inter-dependent Java libraries. However, the construction can be time-consuming. To be practical for alerting, it needs to be demand-driven. Plate et al. \cite{plate2015impact} applied dynamic analysis to check whether the methods that were changed to fix security bugs were executed by a project. However, its effectiveness is limited to the coverage of tests. Then, Ponta et al. \cite{PontaPS2018beyond} extended \cite{plate2015impact} by combining static analysis to partially mitigate the test coverage problem. They are both security-driven, and their alerts report the calls to library APIs that are deleted in the new version. Instead, our alerting system considers non-security bugs and conduct fine-grained change analysis on library APIs by considering their call graphs. Kalra et al. \cite{kalra2016pollux} and Foo et al. \cite{Foo2018ESC} respectively used dynamic and static analysis to find API-breaking changes in libraries, which are useful techniques to improve alerting by locating incompatible API changes. 

\subsection{Evolution and Adaptation} 

A large body of studies have been conducted on API evolution, e.g., impact of refactoring on API breaking \cite{Dig2006AES, Kim2011EIR, Kula2018ESI}, developers' reaction to API evolution \cite{robbes2012developers, hora2015developers, sawant2016reaction}, API stability \cite{raemaekers2012measuring, McDonnell2013ESA, Linares-Vasquez2013ACF}, types of API changes and usages \cite{wu2016exploratory}, adoption of semantic versioning to avoid API breaking \cite{raemaekers2014semantic}, and API breaking in different ecosystems \cite{Bogart2016BAC}. On the other hand, a number of methods have been proposed to adapt to API evolution by change rules written by developers \cite{Chow1996SUA,Balaban2005RSC}, recorded from developers \cite{henkel2005catchup}, derived by similarity matching \cite{xing2007api}, mined from API usage in own libraries \cite{dagenais2009semdiff, dagenais2011recommending}, mined from API usage in projects \cite{schafer2008mining, Nguyen2010GAA}, and identified by a combination of some of these methods \cite{wu2010aura}. Empirical studies have been conducted to compare these methods \cite{cossette2012seeking, wu2015impact}. These methods are a good starting point for automatic library updates.

\subsection{Recommendation and Migration}

 Several approaches have been proposed for library recommendation and migration, e.g., recommending libraries \cite{thung2013automated, Ouni2017SSL}, recommending library APIs \cite{Chan2012SCA, Thung2013ARA}, recommending libraries or library APIs across different programming languages \cite{zheng2011cross, chen2016similartech}, and migration across similar libraries \cite{teyton2012mining, teyton2013automatic, teyton2014study, kabinna2016logging}. However, they do not target the recommendation of and migration between versions of the same library that are useful for automated library updates.


\section{Conclusions}\label{sec:conclusion}

In this paper, we conducted a quantitative and holistic study to characterize usages, updates and risks of third-party libraries in Java open-source projects. Specifically, we quantified the usage and update practices holistically from the perspective of open-source projects and third-party libraries; and we analyzed the risks in terms of bugs for popular third-party libraries.  Our findings provided practical implications to developers and researchers on problems and remedies in maintaining third-party libraries. We also developed a prototype of a bug-driven alerting system for buggy libraries to demonstrate the usefulness of our findings. We released our dataset at \url{https://3rdpartylibs.github.io}  to foster valuable applications and improve the ecosystem of third-party libraries more sustainably. In future, we plan to enhance our bug-driven alerting system by integrating security bugs to have a more complete risk analysis and to developer various techniques to achieve library debloat, library harmonization, automated library update, and automated library bug triggering.


\bibliographystyle{ACM-Reference-Format}
\bibliography{src/reference}

\end{document}